\documentclass{lmcs}
\pdfoutput=1

\usepackage{lastpage}
\lmcsdoi{20}{4}{4}
\lmcsheading{}{\pageref{LastPage}}{}{}%
{Nov.~10,~2023}{Oct.~07,~2024}{}

\keywords{BPMN, Higher-order model transformation, Graph transformation, Model checking, Formalization}
\usepackage[utf8]{inputenc}
\usepackage{booktabs}
\usepackage{footnote}
\usepackage{csquotes} 
\usepackage{threeparttable} 
\makesavenoteenv{tabular}
\makesavenoteenv{table}
\usepackage{courier} 
\usepackage{listings, xcolor}
\definecolor{javared}{rgb}{0.6,0,0} 
\definecolor{javagreen}{rgb}{0.25,0.5,0.35} 
\definecolor{javapurple}{rgb}{0.5,0,0.35} 
\definecolor{javadocblue}{rgb}{0.25,0.35,0.75} 
 
\newlength{\singlelength}
\newlength{\doublelength}
\definecolor{gray}{rgb}{0.4,0.4,0.4}
\definecolor{darkblue}{rgb}{0.0,0.0,0.6}
\definecolor{cyan}{rgb}{0.0,0.6,0.6}

\lstdefinelanguage{XML}
{
  morestring=[b]",
  morestring=[s]{>}{<},
  morecomment=[s]{<?}{?>},
  moredelim=[s][\color{darkblue}]{<}{\ },
  moredelim=[s][\color{darkblue}]{</}{>},
  moredelim=[l][\color{darkblue}]{/>},
  moredelim=[l][\color{darkblue}]{>},
  stringstyle=\color{black},
  identifierstyle=\color{darkblue},
  keywordstyle=\color{cyan},
  morekeywords={id,sourceRef,targetRef, processSnapshot, elementID}
}

\usepackage{hyperref}

\begin{document}

\title[Formalization and analysis of BPMN using graph transformation systems]{A higher-order transformation approach to the formalization and analysis of BPMN using graph transformation systems\rsuper*}
\titlecomment{{\lsuper*}This article is an extended version of \cite{krauterFormalizationAnalysisBPMN2023}}

\author[T.~Kr\"{a}uter]{Tim Kr\"{a}uter\lmcsorcid{0000-0003-1795-0611}}[a]
\author[A.~Rutle]{Adrian Rutle\lmcsorcid{0000-0002-4158-1644}}[a]
\author[H.~K\"{o}nig]{Harald K\"{o}nig\lmcsorcid{0000-0001-6304-6311}}[b,a]
\author[Y.~Lamo]{Yngve Lamo\lmcsorcid{0000-0001-9196-1779}}[a]

\address{Western Norway University of Applied Sciences, Bergen, Norway}
\email{tkra@hvl.no, aru@hvl.no, hkoe@hvl.no, yla@hvl.no}

\address{University of Applied Sciences, FHDW, Hanover, Germany}
\email{harald.koenig@fhdw.de}

\begin{abstract}
  \noindent
The Business Process Modeling Notation (BPMN) is a widely used standard notation for defining intra- and inter-organizational workflows.
However, the informal description of the BPMN execution semantics leads to different interpretations of BPMN elements and difficulties in checking behavioral properties.
In this article, we propose a formalization of the execution semantics of BPMN that, compared to existing approaches, covers more BPMN elements while also facilitating property checking.
Our approach is based on a higher-order transformation from BPMN models to graph transformation systems.
To show the capabilities of our approach, we implemented it as an open-source web-based tool.
\end{abstract}

\maketitle
\section{Introduction} \label{sec:introduction}
In today's fast-paced business environment, organizations with complex workflows require powerful means to accurately map, analyze, and optimize their processes. 
Business Process Modeling Notation (BPMN) \cite{objectmanagementgroupBusinessProcessModel2013} is a widely used standard to define these workflows.
However, the informal description of the BPMN execution semantics leads to different interpretations of BPMN models and difficulties in checking behavioral properties \cite{corradiniFormalApproachAnalysis2021}.
Various studies have shown that business process models suffer from control-flow errors \cite{mendlingEmpiricalStudiesProcess2009}.
Formalizing BPMN can drastically reduce the cost of business process automation by facilitating the detection of errors and optimization potentials in process models already during design time.
For example, general behavioral properties such as \textit{Safeness} and \textit{Soundness} were adapted to BPMN in \cite{corradiniClassificationBPMNCollaborations2018}.
They can uncover control-flow errors in BPMN models leading to deadlocks, dead activities, or other undesirable execution states.
To this end, we propose a formalization that covers nearly all of the BPMN elements used in practice and supports checking behavioral properties to uncover control-flow errors quickly.

\cite{fahlandAnalysisDemandInstantaneous2011} identifies \textit{coverage}, \textit{immediacy}, and \textit{consumability} as the main challenges for applying formal analysis to BPMN models in practice.
In this paper, we focus on \textit{coverage} while we touch upon immediacy (analysis runtime) and consumability (tool usability) when describing the implementation of our approach in \autoref{sec:impl}. 
Many formalizations of BPMN already exist, for example, based on Petri Nets \cite{dijkmanSemanticsAnalysisBusiness2008}, first-order logic \cite{houhouFirstOrderLogicSemantics2019,houhouFirstOrderLogicVerification2022}, and graph transformation \cite{vangorpVisualTokenbasedFormalization2013}.
However, they either do not cover commonly used elements in BPMN~\cite{dijkmanSemanticsAnalysisBusiness2008,houhouFirstOrderLogicSemantics2019,houhouFirstOrderLogicVerification2022} or do not facilitate property checking~\cite{vangorpVisualTokenbasedFormalization2013} (see \autoref{sec:relatedWork}).
Thus, we developed a new formalization that \textbf{(1)} covers all commonly used BPMN elements and \textbf{(2)} supports property checking to uncover control-flow errors.
We can achieve both requirements using advanced graph transformation theory concepts.
Implementing the same coverage, for example, using Petri Nets, is not straightforward.

\begin{figure}[ht]
    \centering
    \includegraphics[width=0.85\textwidth]{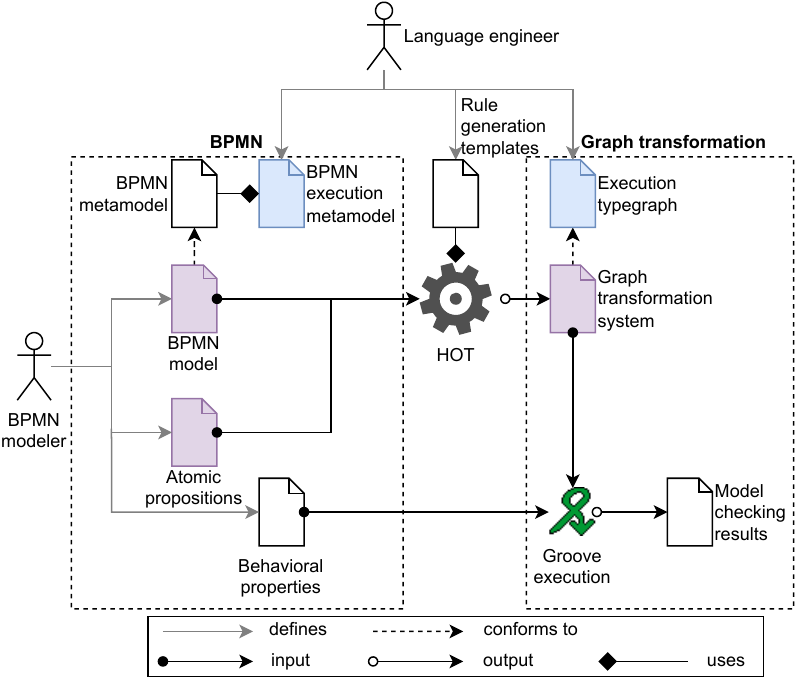}
    \caption{Overview of the approach}
    \label{fig:approach}
\end{figure}

In this article, we consider two fundamental concepts when formalizing the execution semantics of BPMN.
First, \textit{state structure}, i.e., how model instances are represented during execution.
The state structure corresponds to the type graph in Graph Transformation (GT) systems.
Second, \textit{state-changing elements}, i.e., which elements in a model encode state changes.
These elements are implemented using GT rules, which we automatically generate based on a Higher-Order model Transformation (HOT) \cite{tisiUseHigherOrderModel2009} for each specific BPMN model, as shown in \autoref{fig:approach}.
Our HOT defines a formal execution semantics of BPMN, like approaches that formalize BPMN by mapping to Petri Nets or other formalisms \cite{dijkmanSemanticsAnalysisBusiness2008}.

To begin the BPMN modeling process, a modeler first defines the BPMN model. 
The BPMN model may be checked against a predefined list of general behavioral properties, such as safeness and soundness.
Furthermore, the modeler may also define custom behavioral properties specifically defined for the BPMN model.
Custom properties require atomic propositions to describe desirable or undesirable states during execution, which the modeler can define using our concrete syntax based on the BPMN syntax.
The defined BPMN model must adhere to the BPMN metamodel as outlined in the BPMN specification by the Object Management Group \cite{objectmanagementgroupBusinessProcessModel2013}.
The BPMN execution metamodel is defined by language engineers, utilizing the BPMN metamodel as a foundation to create the state structure for executing BPMN models. 

We define a HOT from BPMN models and atomic propositions to GT systems (see purple-colored elements in \autoref{fig:approach}).
We call the transformation \textit{higher-order} since the resulting graph-transformation systems represent model-transformations themselves \cite{tisiUseHigherOrderModel2009}.
The HOT creates a GT system, i.e., GT rules and a start graph for a given BPMN model.
It is defined using rule generation templates, which describe how GT rules should be generated for each state-changing element in BPMN (see \autoref{sec:formalization}).
The obtained GT system conforms to the execution type graph, which corresponds to the BPMN execution metamodel.
In the figure, we have used the same color for artifacts that correspond to each other.
Ultimately, we use Groove to execute the GT system and check the behavioral properties defined earlier.
To facilitate model checking of custom behavioral properties, we create specific GT rules for the corresponding atomic propositions during the HOT.

The overview in \autoref{fig:approach} is divided into two separated parts, denoted by dashed rectangles, to indicate the versatility of the approach as it can be applied to formalize other behavioral languages, such as activity diagrams and state charts \cite{seidlUMLClassroom2015, objectmanagementgroupUnifiedModelingLanguage2017}.
This formalization will require the language engineer to establish a new execution metamodel and a HOT for the new language.
One could even change the \textit{target} of the HOT from GT to a different formalism (term rewriting, Petri Nets, process algebras) if this makes sense for a given behavioral language \cite{krauterBehavioralConsistencyMultimodeling2023}.

This article consists of two main contributions.
First, we introduce a new approach utilizing a HOT to generate GT rules --- instead of providing fixed model-independent GT rules --- to formalize the semantics of a behavioral language.
Second, we apply our approach to BPMN, resulting in a formalization covering most BPMN elements that supports behavioral property checking.
Furthermore, our formalization is implemented as a user-friendly, open-source web-based tool, the \textit{BPMN Analyzer}, which can be used online without needing installation \cite{timkrauterLMCS2024Artifacts2023}.

Our contributions are practical, not theoretical.
We build upon the comprehensive theory and tools available in the GT research field.
Concretely, we utilize the single-pushout (SPO) approach with negative application conditions (NAC) \cite{ehrigALGEBRAICAPPROACHESGRAPH1997}, as implemented in Groove \cite{rensinkGROOVESimulatorTool2004}.
In addition, we utilize \textit{nested rules} with quantification to make parts of a rule repeatedly applicable or optional \cite{rensinkNestedQuantificationGraph2006,rensinkHowMuchAre2017}.
Moreover, we utilize the NACs to implement more intricate parts in the BPMN execution semantics, such as the termination of processes.
Formal definitions of SPO rules, their application, and the corresponding extensions of the theory (NACs, nested rules) are well-known, see \cite{ehrigALGEBRAICAPPROACHESGRAPH1997,rensinkNestedQuantificationGraph2006}.
We do not repeat them and instead focus on our practical contribution.

This article extends \cite{krauterFormalizationAnalysisBPMN2023} as follows.
(i) We explain many more BPMN elements, which are covered by our approach (see elements highlighted in blue in \autoref{fig:bpmnelementsOverview}).
(ii) We enhance the explanation of the custom properties in \autoref{sec:modelChecking} by using an order handling process to illustrate use cases for these properties.
(iii) We detail the extensively improved BPMN analyzer tool in \autoref{sec:impl} in which modelers can use our new atomic proposition editor.
(iv) We test the scalability of our approach with 300 synthetically generated BPMN models of increasing size in \autoref{sec:impl}. 

\textbf{Outline} The remainder of this article is structured as follows.
First, we describe the BPMN semantics formalization using the HOT (\autoref{sec:formalization}) before explaining how this can be utilized for model checking general BPMN and custom properties (\autoref{sec:modelChecking}).
Then, we detail the BPMN Analyzer, which implements our approach in \autoref{sec:impl} and describe how we tested its performance and scalability.
Finally, we discuss related work regarding BPMN element coverage in \autoref{sec:relatedWork} and conclude in \autoref{sec:conclusion}.

\section{BPMN syntax \& BPMN semantics formalization} \label{sec:formalization}

\begin{figure}[ht]
  \centering
  \includegraphics[width=0.8\linewidth]{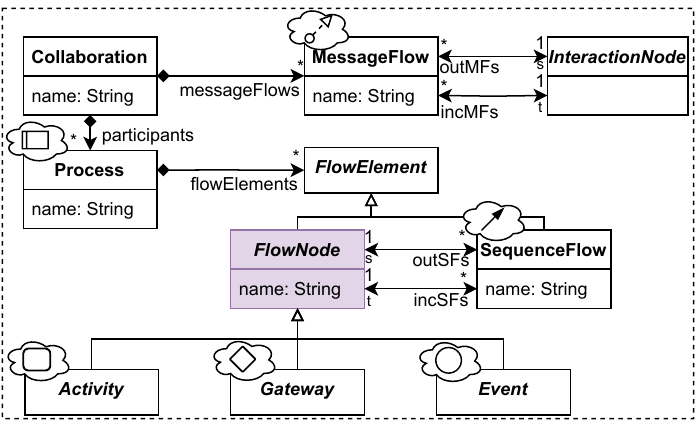}
  \caption{Simplified excerpt of the BPMN metamodel \cite{objectmanagementgroupBusinessProcessModel2013}}
  \label{fig:bpmnMetamodel}
\end{figure}

\autoref{fig:bpmnMetamodel} depicts the structure of BPMN models with the corresponding concrete syntax BPMN symbols contained in clouds.
A BPMN model is represented by a \textsf{Collaboration} that has participant \textsf{Process}es and \textsf{MessageFlow}s between \textsf{InteractionNode}s.
Each participant is a \textsf{Process} containing \textsf{FlowElement}s.
A \textsf{FlowElement} is either a \textsf{FlowNode} or \textsf{SequenceFlow}.
A \textsf{FlowNode} is either an \textsf{Activity}, a \textsf{Gateway}, or an \textsf{Event} and can be connected to other \textsf{FlodNode}s using \textsf{SequenceFlow}s.
Many types of activities, gateways, and events exist, such as call activities, parallel gateways, and start events.
Activities represent certain tasks to be carried out during a process, while events may happen during the execution of these tasks.
Furthermore, gateways model conditions, parallelizations, and synchronizations \cite{freundRealLifeBPMNUsing2019}.

\autoref{fig:bpmnMetamodel} is a simplified excerpt of the BPMN metamodel described in the BPMN specification \cite{objectmanagementgroupBusinessProcessModel2013}.
Each class depicted in \autoref{fig:bpmnMetamodel} maintains consistent naming with the BPMN metamodel, and all associations are directly found in \cite{objectmanagementgroupBusinessProcessModel2013} or simplified, i.e., given by compositions of multiple existing associations.

\begin{figure}[ht]
    \centering
    \includegraphics[width=0.99\textwidth]{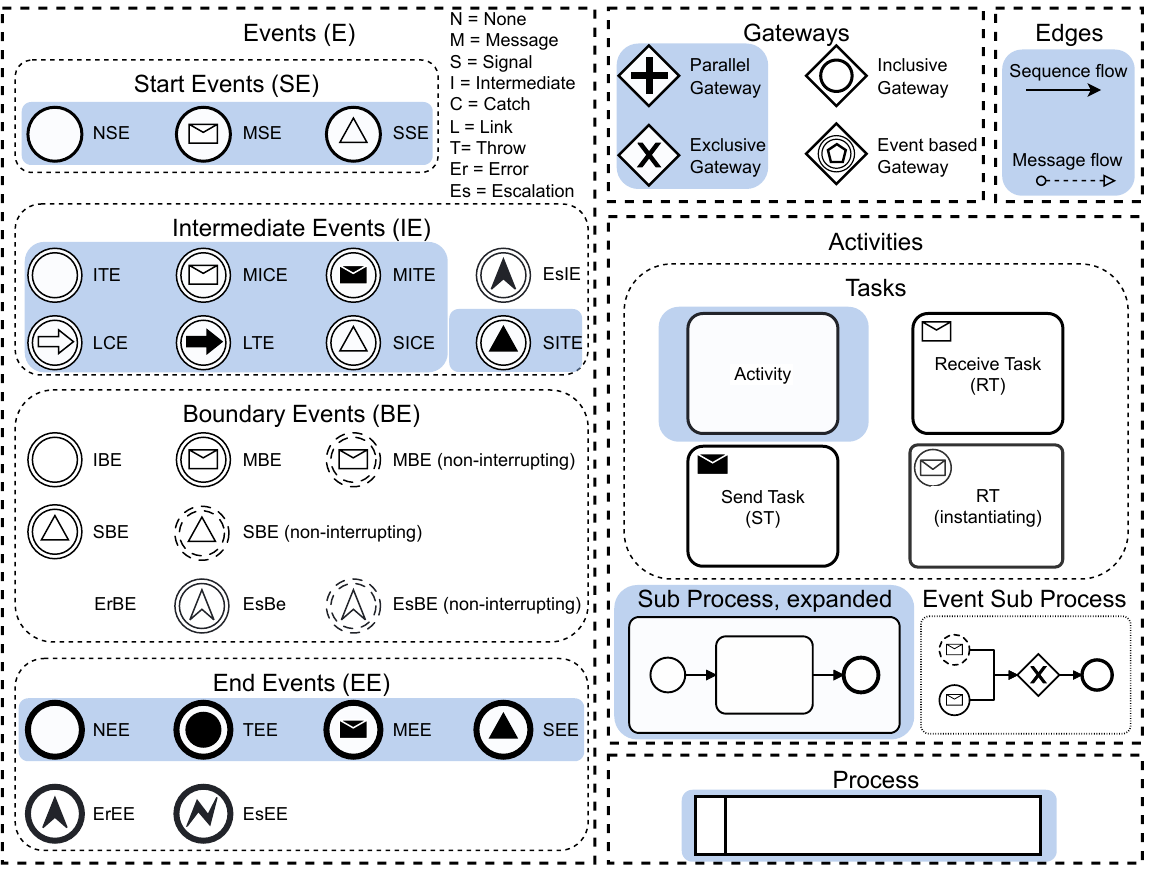}
    \caption{Overview of the supported BPMN elements (structure adapted from \cite{houhouFirstOrderLogicVerification2022})}
    \label{fig:bpmnelementsOverview}
\end{figure}

Our approach supports all the BPMN elements depicted in \autoref{fig:bpmnelementsOverview}.
These BPMN elements are divided into \textsf{Events}, \textsf{Gateways}, \textsf{Activities}, and \textsf{Edges}.
\textsf{Events} and \textsf{Activities} are further divided into subgroups.
An extensive overview of BPMN and its elements can be found in \cite{freundRealLifeBPMNUsing2019}.
Although all these elements have been implemented and tested (see \cite{timkrauterLMCS2024Artifacts2023}), we only explain the realization of the elements marked with a blue background due to space limitations.
In the following, first, we define the BPMN execution metamodel to represent the BPMN state structure, and then we explain our formalization of the elements in \autoref{fig:bpmnelementsOverview}.

\subsection{BPMN execution metamodel}

The BPMN execution semantics is described using the concept of \textit{tokens} \cite{objectmanagementgroupBusinessProcessModel2013, freundRealLifeBPMNUsing2019}, which can be located at sequence flows and specific flow nodes.
Tokens are consumed and created by flow nodes according to the flow node's type and the connected sequence flows.
The \textsf{FlowNode} is colored purple in \autoref{fig:bpmnMetamodel} since it represents the \textit{state-changing elements} of BPMN, as described in \autoref{sec:formalization}.
In our formalization of BPMN, we follow this token-based representation of the execution semantics.

\begin{figure}[ht]
  \centering
  \includegraphics[width=1\linewidth]{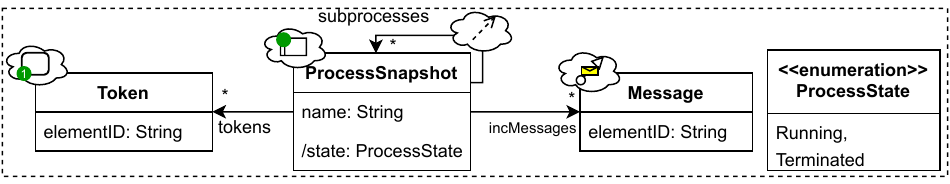}
  \caption{BPMN execution metamodel}
  \label{fig:bpmnExecutionMetamodel}
\end{figure}

To describe processes holding tokens during execution, we define the execution metamodel shown in \autoref{fig:bpmnExecutionMetamodel}, depicted as a UML class diagram.
The first task of a language engineer in our approach is to define the execution metamodel (see \autoref{fig:approach}).

The BPMN execution metamodel was not created by extending the BPMN metamodel and adding missing concepts such as tokens and messages.
We created a minimal execution model that only contains concepts needed during execution.
This is only possible since the HOT generates our rules for each specific BPMN model such that the structure of each model is already implicitly encoded in the rules.
This design choice leads to smaller states in the GT system compared to an execution metamodel that extends the BPMN metamodel.

In \autoref{fig:bpmnExecutionMetamodel}, we use \textsf{ProcessSnapshot} to denote a running BPMN process with a specific token distribution that describes one state in the history of the process execution.
Every \textsf{ProcessSnapshot} has a set of \textsf{tokens}, incoming \textsf{messages}, and \textsf{subprocesses}.
A \textsf{ProcessSnapshot} has the state \textsf{Terminated} if it has no \textsf{tokens} or \textsf{subprocesses}.
Otherwise, it has the state \textsf{Running}.
A \textsf{Token} has an \textsf{elementID}, which points to the BPMN \textsf{Activity} or the \textsf{SequenceFlow} at which it is located.
A \textsf{Message} has an \textsf{elementID} pointing to a \textsf{MessageFlow}.
To concisely depict graphs conforming to this type graph, we introduce a concrete syntax in the clouds attached to the elements.
Our concrete syntax extends the BPMN syntax by adding process snapshots, subprocess relations, tokens, and messages.
Tokens are represented as colored circles drawn at their specified positions in a model.
In addition, we use colored circles at the top left of the bounding box, representing instances of the BPMN \textsf{Process}; these circles represent process snapshots.
The token's color must match the color of the process snapshot holding the token.
The concrete syntax was inspired by the bpmn-js-token-simulation \cite{camundaservicesgmbhBpmnjsTokenSimulation2023}.

The execution metamodel in \autoref{fig:bpmnExecutionMetamodel} is a UML class diagram without operations, which can be seen as an attributed type graph \cite{heckelGraphTransformationSoftware2020}.
We keep the execution metamodel and the execution type graph separate (see \autoref{fig:approach}) because the execution metamodel should be independent of the formalism used to define the execution semantics.
One can reuse the execution metamodel when changing the formalism or concrete tool implementing the formalism (in our case, Groove) by adjusting how the execution metamodel is transformed.
Using the execution metamodel as the type graph, we can now define how the start graph and GT rules for the different BPMN elements are created.

\begin{figure}[ht]
    \centering
    \includegraphics[width=0.85\textwidth]{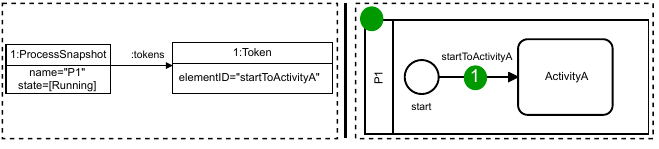}
    \caption{Example start graph in abstract (left) and concrete syntax (right)}
    \label{fig:startGraph}
\end{figure}

Since our approach is based on a HOT from BPMN to GT systems, we generate a \textit{start graph} and \textit{GT rules} for each given BPMN model (see \autoref{fig:approach}).
Generating the start graph for a BPMN model is straightforward.
First, for each process in the BPMN model, we generate a process snapshot if the process contains a \textit{None Start Event} (NSE).
An NSE describes a start event without a trigger (none).
Then, for each NSE, we add one token to each outgoing sequence flow.
An example of a start graph is shown in \autoref{fig:startGraph} using abstract and concrete syntax.

The HOT generates one or more GT rules for each \textsf{FlowNode}, i.e., state-changing element in a BPMN model.
To better understand the transformation process, we will begin by presenting example results, namely the generated rules for an activity.
Following this, we will explain how our HOT creates these rules and rules for the other elements in \autoref{fig:bpmnelementsOverview}.

\begin{figure}[ht]
    \centering
  \includegraphics[width=1\textwidth]{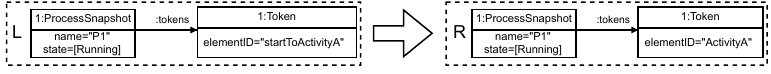}
  \caption{Example GT rule to start an activity (abstract syntax)}  \label{fig:gtRuleAbstract}
\end{figure}

\autoref{fig:gtRuleAbstract} depicts an example GT rule ($L \to R$) to start an activity in abstract syntax.
The rule is straightforward: it moves a token from the incoming sequence flow to the activity.

\begin{figure}[ht]
    \centering
  \includegraphics[width=0.7\textwidth]{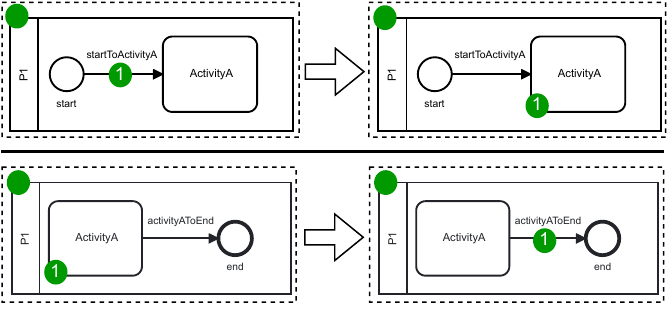}
  \caption{Example GT rules to start (top) and terminate (bottom) an activity}
  \label{fig:gtRuleConcrete}
\end{figure}

For the rest of the article, we will depict all rules in the concrete syntax introduced earlier.
The rule from \autoref{fig:gtRuleAbstract} depicted in concrete syntax is shown on the top in \autoref{fig:gtRuleConcrete}.
The rule on the bottom in \autoref{fig:gtRuleConcrete} implements the termination of an activity, which will move one token from the activity to the outgoing sequence flow.

In the following subsections, we use our concrete syntax to define how our HOT generates these rules and rules for other flow nodes.
Elements of the HOT are depicted using rule generation templates that show how specific rules are created for various flow nodes.
Defining the rule generation templates and, thus, the HOT from BPMN to GT systems is the second task of the language engineer in our approach (see \autoref{fig:approach}).

\subsection{Process instantiation and termination} \label{subsec:instAndTermination}

Start events do not need GT rules since the generated start graph of the GT system will contain a token for each outgoing sequence flow of an NSE.
Other types of start events are triggered in corresponding throw event rules.

\begin{figure}[ht]
    \centering
    \includegraphics[width=.8\textwidth]{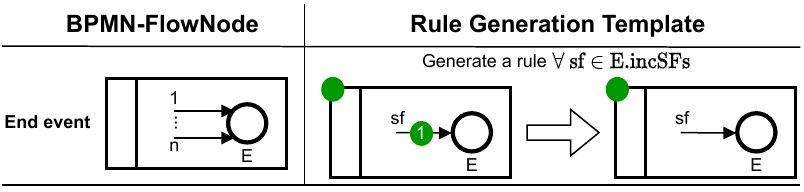}
    \caption{Rule generation template for \textit{None End Events}}
    \label{fig:endTemplate}
\end{figure}

\autoref{fig:endTemplate} depicts the rule generation template for \textit{None End Events} (\textsf{NEE}s in \autoref{fig:bpmnelementsOverview}).
All rule generation templates show a state-changing element (\textsf{FlowNode}) with surrounding flows in the left column and the applicable rule generation in the right column.
The left column shows instances of the BPMN metamodel (\autoref{fig:bpmnMetamodel}), and the right column shows the generated rules typed by the BPMN execution metamodel (see \autoref{fig:bpmnExecutionMetamodel}).
If more than one rule is generated from a \textsf{FlowNode}, an expression defines how each rule is generated.
For example, the expression $\forall \text{sf} \in \text{E.incSFs}$ for the rule generation template of end events (see \autoref{fig:endTemplate}) generates one rule for each incoming sequence flow \textit{sf} of the end event \textit{E}.
We use ``.'' in expressions to navigate along the associations of the BPMN metamodel shown in \autoref{fig:bpmnMetamodel}.
In the example, \textsf{E.incSFs} means following all \textsf{incSFs} links for a \textsf{FlowNode} object, resulting in a set of \textsf{SequenceFlow} objects.

For example, if a BPMN model contains one \textsf{NEE} with two incoming sequence flows, the HOT will generate two GT rules as defined in the rule generation template in \autoref{fig:endTemplate}.
Thus, each rule generation template defines part of the HOT, and all rule generation templates together represent the HOT, which can be applied to a given BPMN model to generate concrete rules as specified in each template.

\begin{figure}[ht]
    \centering
    \includegraphics[width=.4\textwidth]{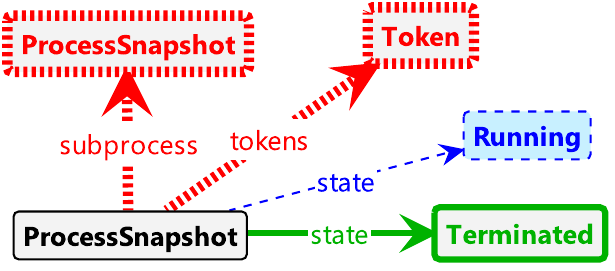}
    \caption{Termination rule in Groove}
    \label{fig:terminationRule}
\end{figure}

The generated end event rules delete tokens individually for each incoming sequence flow.
However, they do not terminate processes.
Process termination is implemented with a generic rule---independent of the input BPMN model---which applies to all process snapshots.
The termination rule in \autoref{fig:terminationRule} is automatically generated once during the HOT.
The rule changes the state of the process snapshot from running to terminated if it has neither tokens nor subprocesses.
We use Groove Syntax instead of our concrete syntax since it is a special, more complex rule, and we do not want the concrete syntax to become too complex.

The Groove syntax is the following.
The thin black elements in \autoref{fig:terminationRule} need to be present and will be preserved during transformation, while the dashed blue elements need to be present but will be removed.
Furthermore, the fat green elements will be created, and the dashed fat red elements represent the NACs, whose presence prevents the rule from being applied.

\subsection{Activities \& Subprocesses}
Activities represent work to be performed within a BPMN process, while subprocesses group parts of a BPMN model together, allowing for reusability and separation of concerns \cite{objectmanagementgroupBusinessProcessModel2013}.

\begin{figure}[ht]
    \centering
    \includegraphics[width=1\textwidth]{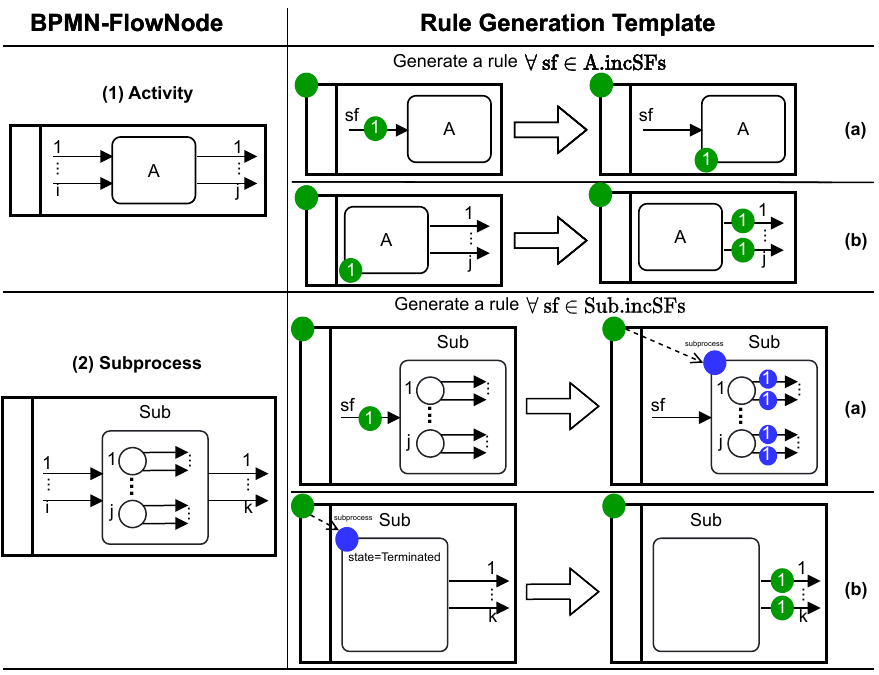}
    \caption{Rule generation template for activities and subprocesses}
    \label{fig:activityTemplates}
\end{figure}

\autoref{fig:activityTemplates} depicts the rule generation templates for activities and subprocesses (see \autoref{fig:bpmnelementsOverview}).
Activity execution is divided into two steps implemented in parts \textbf{(a)} and \textbf{(b)} in the rule generation template \textbf{(1)}.
Part \textbf{(a)} generates one rule for each incoming sequence flow to start the activity.
An activity can be started using a token positioned at any of its incoming sequence flows.
This part generates the sample rule on the left of \autoref{fig:gtRuleConcrete}.
Having multiple incoming or outgoing sequence flows for a flow node is considered bad practice since the implicitly encoded gateways should be explicit to avoid confusion.
Our formalization still supports those models not to force modelers to rewrite them, but we recommend using static analyzers to avoid such models \cite{camundaservicesgmbhBpmnlint2023}.

Part \textbf{(b)} generates one rule that terminates the activity.
It deletes a token at the activity and adds one at each outgoing sequence flow.
This implicitly encodes a parallel gateway (see \autoref{fig:gatewayTemplates}) but should be avoided, as described earlier. 

Subprocess execution is like activity execution.
Part \textbf{(a)} of the template generates one rule for each incoming sequence flow.
The rule deletes an incoming token and adds a process snapshot representing a subprocess. 
The created process snapshot is represented with a colored circle on the top left corner of the subprocess with a token at each outgoing sequence flow of its start events (similar to start graph generation).
There is a \textit{subprocess} link between the process snapshots to depict the \textsf{subprocesses} relation in \autoref{fig:bpmnExecutionMetamodel}.
If the subprocess has no start events, a token will be added to every activity and gateway with no incoming sequence flows.

Part \textbf{(b)} of the template generates one rule to delete a terminated process snapshot and adds tokens at each outgoing sequence flow.
Subprocesses are terminated by the termination rule (see section \ref{subsec:instAndTermination}).

\subsection{Gateways}
Parallel gateways represent forking and joining in the sequence flow.
Exclusive gateways represent exclusive choices and merges in the sequence flow \cite{objectmanagementgroupBusinessProcessModel2013}.

\begin{figure}[ht]
    \centering
    \includegraphics[width=0.85\textwidth]{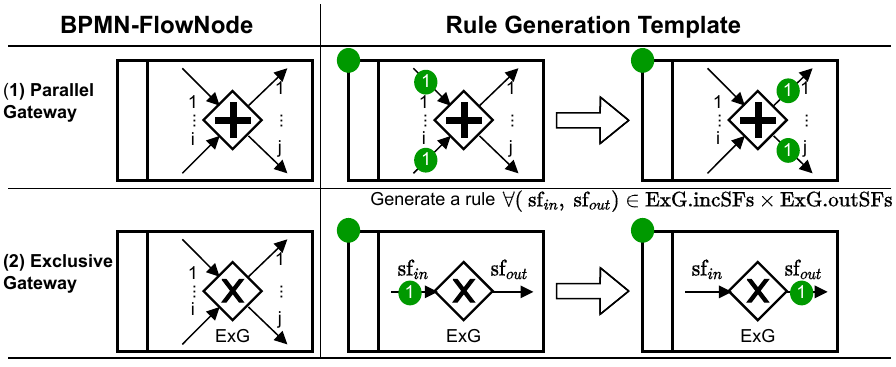}
    \caption{Rule generation template for gateways}
    \label{fig:gatewayTemplates}
\end{figure}

\autoref{fig:gatewayTemplates} depicts the rule generation templates for parallel and exclusive gateways (see \autoref{fig:bpmnelementsOverview}).
A parallel gateway can synchronize and fork the control flow simultaneously.
Thus, one rule is generated that deletes one token from each incoming sequence flow and adds one to each outgoing sequence flow.

Exclusive Gateways are triggered by exactly one incoming sequence flow, and exactly one outgoing sequence flow will be triggered as a result.
In practice, boolean conditions using data attached to the process are attached to exclusive gateways that decide which outgoing sequence flow to follow.
We do not model data flow in our formalization and instead allow each outgoing sequence flow to be explored. 
Thus, one rule must be generated for every combination of incoming and outgoing sequence flows.
However, the resulting rule is simple since it only deletes a token from an incoming sequence flow and adds one to an outgoing sequence flow.

\subsection{Message Events}
\begin{figure}[ht]
    \centering
    \includegraphics[width=1\textwidth]{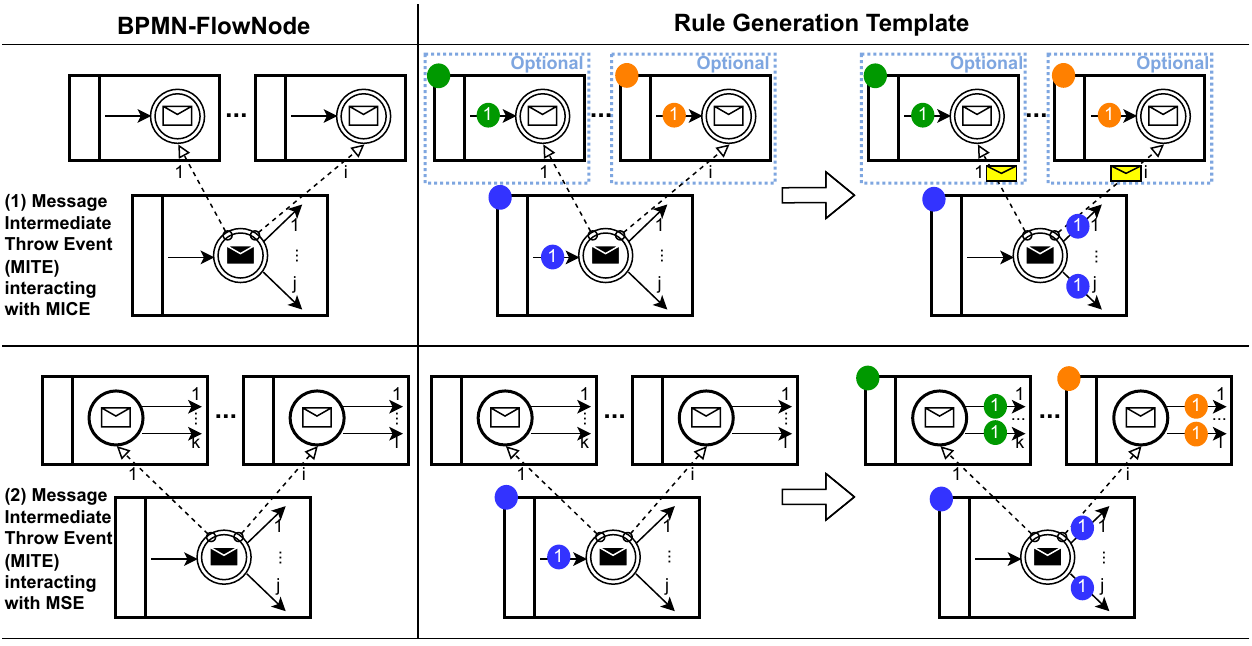}
    \caption{Rule generation templates for MITEs interacting with MICEs \textbf{(1)} and MSEs \textbf{(2)}}
    \label{fig:messageThrowEventTemplates}
\end{figure}

Message events are events directed at a single recipient.
Thus, they are unicast compared to broadcast signal events discussed later in \autoref{subsec:signalEvents}.
\autoref{fig:messageThrowEventTemplates} depicts the rule generation templates for \textit{Message Intermediate Throw Events} (\textsf{MITE} in \autoref{fig:bpmnelementsOverview}).
Rule generation template \textbf{(1)} describes how MITEs interact with \textit{Message Intermediate Catch Events} (MICEs).
A MITE deletes an incoming token and adds one at each outgoing sequence flow.
In addition, it sends one message to each process by adding it to the incoming messages of the process.
However, sending each message is optional, meaning that if a process is not ready to consume a message immediately, the message is not added.
A process can consume a message if its MICE has at least one token at an incoming sequence flow (see rule template (1) in \autoref{fig:messageThrowEventTemplates}).
We implement optional message sending using nested rules with quantification.
Concretely, we use an optional existential quantifier \cite{rensinkNestedQuantificationGraph2006} (see dotted rectangle labeled \textsf{Optional} in \autoref{fig:messageThrowEventTemplates}) to send a message only if the receiving process is ready to consume it.

Rule generation template \textbf{(2)} describes how MITEs interact with \textit{Message Start Events} (MSEs).
For each MSE, a new process snapshot is created with tokens located at its outgoing sequence flows.
We split the interaction of MITEs with MICEs and MSEs into two rule templates for better understanding.
However, a MITE might interact with MICEs and MSEs simultaneously.
Thus, our HOT implements a merge of both templates. 
\textit{Message End Events} (\textsf{MEE}) behave similarly to MITEs but only delete incoming tokens and do not add outgoing tokens.

\begin{figure}[ht]
    \centering
    \includegraphics[width=0.8\textwidth]{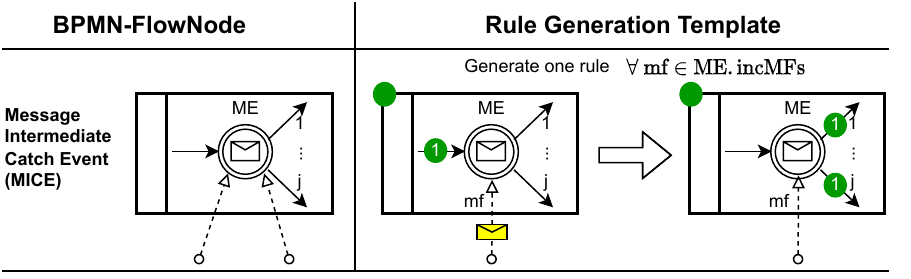}
    \caption{Rule generation templates for MICEs}
    \label{fig:messageCatchEventTemplates}
\end{figure}

The rule generation template in \autoref{fig:messageCatchEventTemplates} shows the behavior of MICE (see \textsf{MICE} in \autoref{fig:bpmnelementsOverview}).
To trigger a MICE, only one message at an incoming \textit{message flow} is needed.
Thus, one rule is generated for each incoming \textit{message flow}.
The rule template shows that MICEs delete one message and one token, as well as add a token at each outgoing sequence flow.

\subsection{Link Events} \label{subsec:signalEvents}
\begin{figure}[ht]
    \centering
    \includegraphics[width=0.85\textwidth]{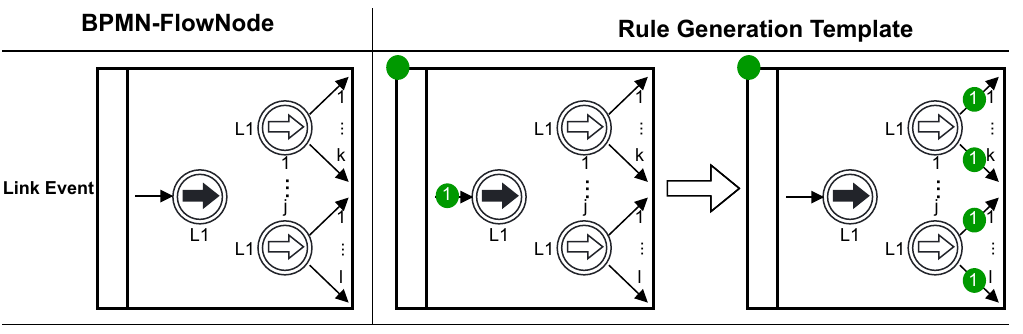}
    \caption{Rule generation template for LTEs interacting with LCEs}
    \label{fig:linkEventTemplates}
\end{figure}

Link events are similar to \enquote{Go To} statements since they move tokens from link throw events to link catch events in the same process level (cannot link to subprocesses).
They are meant to avoid long sequence flows and connect BPMN models spanning multiple pages but can also be used to create loops due to their \enquote{Go To} nature \cite{objectmanagementgroupBusinessProcessModel2013}.
\autoref{fig:linkEventTemplates} depicts the rule generation template for \textit{Link Throw Events} (LTEs), see \textsf{LTE} in \autoref{fig:bpmnelementsOverview}.
It shows how LTEs interact with \textit{Link Catch Events} (LCEs).

Each rule deletes a token at that sequence flow and adds tokens to all outgoing sequence flows of matching LTEs.
An LTE matches an LCE if they have the same name or event definition (see \cite{objectmanagementgroupBusinessProcessModel2013}).
Our HOT automatically finds matching LTEs during transformation and then applies the rule template shown in \autoref{fig:linkEventTemplates}.

\subsection{Signal Events}
Each signal event is assigned a signal name.
Signal throw events \textit{broadcast} to all signal catch events with the same signal name.
Signal broadcasts have a global scope, i.e., they can communicate across process levels and pools \cite{objectmanagementgroupBusinessProcessModel2013}.

\begin{figure}[ht]
    \centering
    \includegraphics[width=1\textwidth]{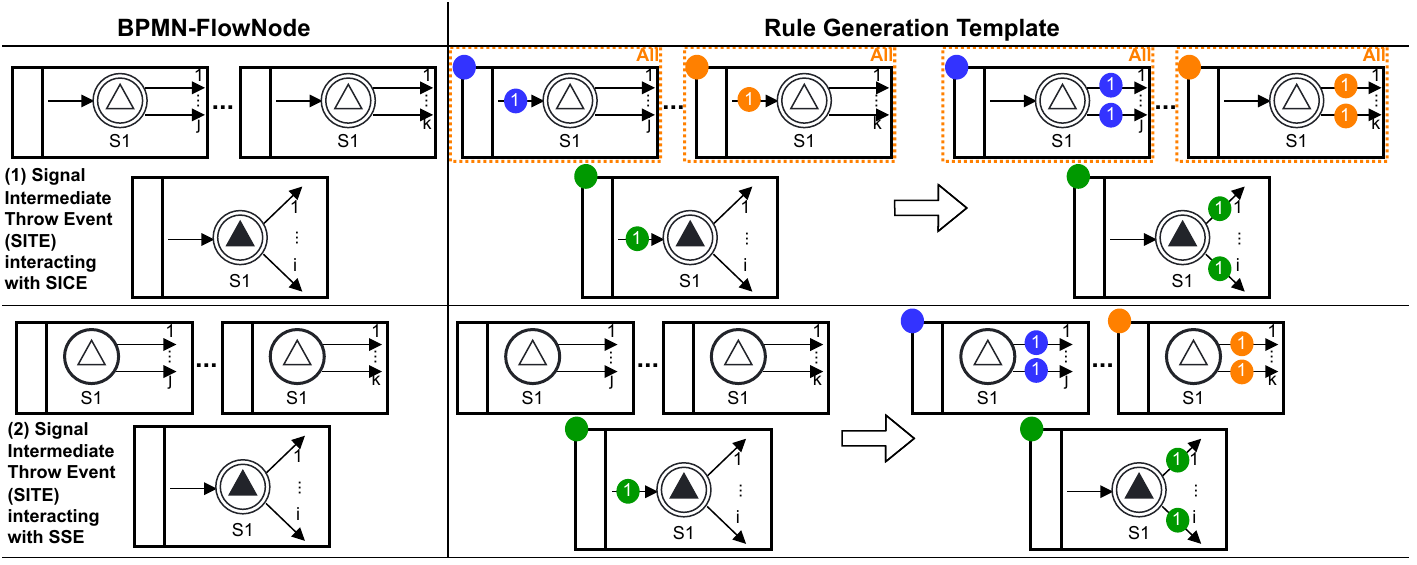}
    \caption{Rule generation templates for SITEs interacting with SICE \textbf{(1)} and SSE \textbf{(2)}}
    \label{fig:signalEventTemplates}
\end{figure}

\autoref{fig:signalEventTemplates} depicts the rule generation template for \textit{Signal Intermediate Throw Events} which interact with \textit{Signal Intermediate Catch Events} and \textit{Signal Start Events} (\textsf{SITE}, \textsf{SICE}, and \textsf{SSE} in \autoref{fig:bpmnelementsOverview}).
\textit{Signal End Events} (\textsf{SEE}) behave similarly to SITEs but only consume incoming tokens and do not add outgoing tokens.

Rule generation template \textbf{(1)} describes how SITEs interact with SICE.
Like other intermediate events, the incoming token is consumed while one token is added for each outgoing sequence flow.
Due to its broadcast semantics, a SITE interacts with all matching SICEs with an incoming token.
A SITE and SICE match if they have the same signal name.
In our templates, we assume that the signal name is the SITE/SICE name.
For each matching SICE, a universally quantified nested rule consumes the incoming token and adds a token for each outgoing sequence flow.
We use a universal quantifier (\textsf{All} in \autoref{fig:signalEventTemplates}) since one process snapshot might have multiple tokens waiting before a SICE.
Then, a SITE should trigger this SICE multiple times.

Rule generation template \textbf{(2)} describes how SITEs interact with SSEs.
Analogous to MITEs and MSEs, new process snapshots with tokens at the outgoing sequence flows of the SSEs are added for each matching SSE.
Each matching SSE is only triggered once, meaning we do not need any quantified nested rules.
We split the interaction of SITEs with SICEs and SSEs into two rule templates for better understanding.
However, a SITE might interact with SICEs and SSEs simultaneously.
Thus, our HOT implements a merge of both templates.

\subsection{Terminate Events}
\begin{figure}[ht]
    \centering
    \includegraphics[width=0.8\textwidth]{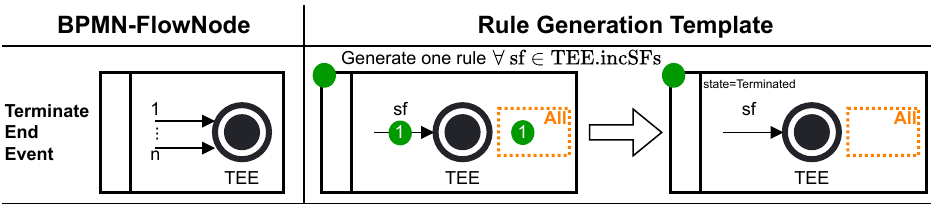}
    \caption{Rule generation template for terminate end events}
    \label{fig:terminateEventTemplate}
\end{figure}

A \textit{Terminate End Event} (TEE) abnormally terminates the running process \cite{objectmanagementgroupBusinessProcessModel2013}, meaning the process changes its state to terminated, and all its tokens are consumed.
\autoref{fig:terminateEventTemplate} depicts the rule generation template for TEEs (see \textsf{TEE} in \autoref{fig:bpmnelementsOverview}).

One rule is generated for each incoming sequence flow of a TEE.
The rule consumes the incoming token, similar to the rules for end events, but also changes the process snapshot state to \textsf{Terminated}.
In addition, the rule deletes all other tokens of the process snapshot using a universally quantified nested rule (see dotted rectangle labeled \textbf{All} in \autoref{fig:terminateEventTemplate}).
Terminating a process must also terminate its subprocesses, which is not shown in the rule template in \autoref{fig:terminateEventTemplate} for brevity; it is described in our wiki \cite{timkrauterLMCS2024Artifacts2023}.

\section{Model checking BPMN} \label{sec:modelChecking}

Model checking ---and verification in general--- of BPMN models is necessary to ensure the correctness and reliability of business processes, which ultimately leads to increased efficiency, reduced costs, and user satisfaction.
Using our formalization approach, BPMN models may be verified against behavioral properties ---both general and custom--- by utilizing the generated GT system (see \autoref{fig:approach}).
These behavioral properties are defined using temporal logic, such as Computation Tree Logic (CTL) and Linear Temporal Logic (LTL) \cite{baierPrinciplesModelChecking2008,clarkeHandbookModelChecking2018}.
As mentioned in \autoref{sec:introduction}, modelers may use our approach to specify custom properties consisting of atomic propositions and operators from CTL/LTL.
The atomic propositions are transformed to graph conditions by the HOT.
A graph condition is a GT rule which does not delete or add elements.
A proposition holds in a given state if a match of the graph condition exists in the graph representing the state \cite{kastenbergModelCheckingDynamic2006}.

We differentiate between two types of behavioral properties: \textit{general BPMN properties} defined for all BPMN models and \textit{custom properties} tailored towards a particular BPMN model.
We do not consider structural properties (like conformance to BPMN syntax) since they can be checked using a standard modeling tool without implementing execution semantics.
We will now give an example of predefined general BPMN properties and show how our approach can check them.
Then, we describe how custom properties can be defined and checked.

\subsection{General BPMN properties}
\textit{Safeness} and \textit{Soundness} properties are defined for BPMN in \cite{corradiniClassificationBPMNCollaborations2018}.
A BPMN model is \textit{safe} if, during its execution, at most one token occurs along the same sequence flow \cite{corradiniClassificationBPMNCollaborations2018}.
Soundness is further decomposed into (i) \textit{Option to complete}: any running process instance must eventually complete, (ii) \textit{Proper completion}: after completion, each token of the process instance must be consumed by a different end event, as well as (iii) \textit{No dead activities}: each activity can be executed in at least one process instance \cite{corradiniClassificationBPMNCollaborations2018}.
Process completion is synonymous with process termination.
In the following, we will describe how to implement the \textit{Safeness} and \textit{Option to complete} using CTL, as well as \textit{Proper completion} and \textit{No dead activities} by analyzing the GT system's state space.

\begin{figure}[ht]
    \centering
    \includegraphics[width=0.5\textwidth]{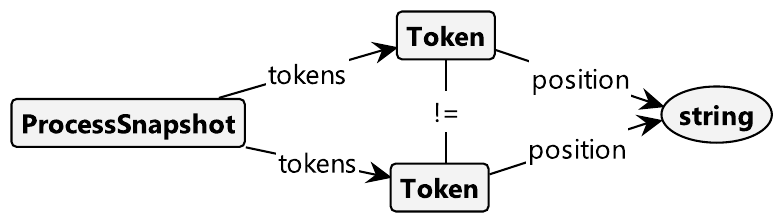}
    \caption{The atomic proposition \textsf{Unsafe} as a Groove graph condition.}
    \label{fig:unsafe}
\end{figure}

We specify \textit{Safeness} as the following CTL property:
\begin{equation} \label{eq:safeness}
  AG(\neg \,\text{Unsafe})
\end{equation}
\vskip.1\baselineskip
The path quantifier $A$ means the following proposition $G(\neg \,\text{Unsafe})$ should hold for \textit{all} paths starting from the current state.
The temporal operator $G$ means the following proposition $\neg \,\text{Unsafe}$ should hold at all states in the future \cite{clarkeHandbookModelChecking2018}.
More detailed information about CTL can be found in \cite{clarkeHandbookModelChecking2018, baierPrinciplesModelChecking2008}.
Combining the path quantifier $A$ and temporal operator $G$ in \eqref{eq:safeness} means $\neg \,\text{Unsafe}$ should hold for all states in all paths starting from the initial state.
Thus, \eqref{eq:safeness} describes that a state labeled as \textsf{Unsafe} should not be reachable.
The atomic proposition \textsf{Unsafe} is true if two tokens of one process snapshot point to the same sequence flow.
Atomic propositions are either fulfilled
\autoref{fig:unsafe} shows how \textsf{Unsafe} is represented as a graph condition in Groove.

\begin{figure}[ht]
    \centering
    \includegraphics[width=0.35\textwidth]{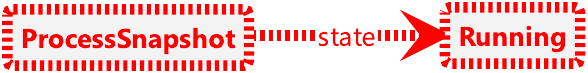}
    \caption{The atomic proposition \textsf{AllTerminated} as a Groove graph condition.}
    \label{fig:allTerminated}
\end{figure}

\textit{Option to complete} is specified using the following CTL property:
\begin{equation} \label{eq:optionToComplete}
  AF(\text{AllTerminated}) 
\end{equation}
\vskip.1\baselineskip
The temporal operator $F$ means the following proposition \textsf{AllTerminated} should hold in some state in the future \cite{clarkeHandbookModelChecking2018}.
Thus, \eqref{eq:optionToComplete} describes that a state labeled as \textsf{AllTerminated} should be reached for all paths starting from the initial state.
The atomic proposition \textsf{AllTerminated} is true if there exists no process snapshot in the state \textsf{Running}, i.e., all process snapshots are \textsf{Terminated}.
\autoref{fig:allTerminated} shows how \textsf{Terminated} is represented as a graph condition in Groove.

Checking the properties \textit{Safeness} and \textit{Option to complete} is implemented by checking the CTL properties above using Groove \cite{kastenbergModelCheckingDynamic2006, rensinkExplicitStateModel2008}.
The property \textit{Proper Completion} is implemented by checking the GT system's state space for two executions of an end event in the same path.
Similarly, \textit{No dead activities} is implemented by analyzing the GT system's state space to see if each activity has been executed at least once \cite{timkrauterLMCS2024Artifacts2023}.

\subsection{Custom properties} \label{subsec:customProperties}

\begin{figure}[ht]
    \centering
    \includegraphics[width=0.55\textwidth]{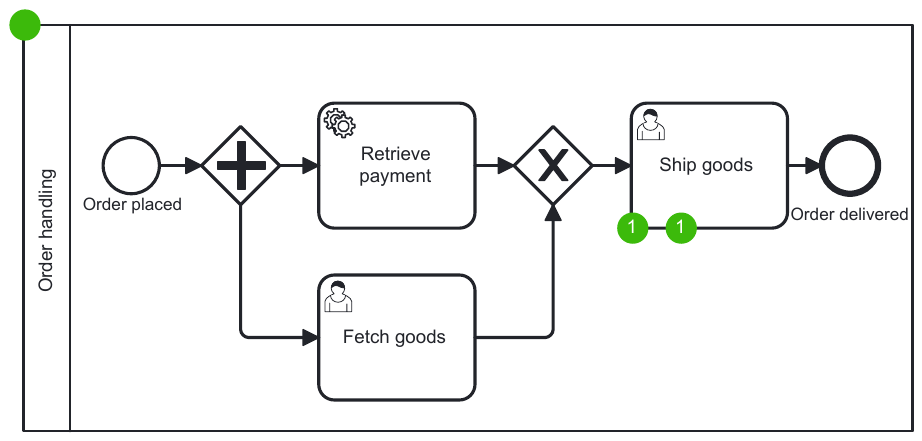}
    \caption{Atomic proposition \textit{shipGoodsTwice} defining shipping goods twice.}
    \label{fig:shippedTwiceProposition}
\end{figure}

\begin{figure}[ht]
    \centering
    \includegraphics[width=0.5\textwidth]{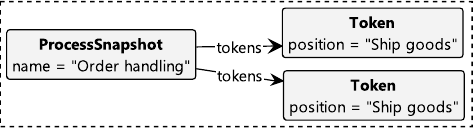}
    \caption{Generated Groove graph condition describing the atomic proposition in \autoref{fig:shippedTwiceProposition}.}
    \label{fig:shippedTwiceGroove}
\end{figure}

To make model checking user-friendly, we enable modelers to define atomic propositions using the concrete syntax of the extended BPMN execution metamodel introduced in \autoref{fig:bpmnExecutionMetamodel} (see \autoref{fig:shippedTwiceProposition}).
An atomic proposition is defined as a process snapshot with a token distribution, which we can automatically convert to a graph condition in Groove (see \autoref{fig:shippedTwiceGroove}).
Recall that graph conditions are GT rules that do not add or delete elements.
Atomic propositions may be connected by CTL operators to create temporal formulas that should hold in the given BPMN model. 
Furthermore, modelers may forbid certain states in the BPMN model by specifying that a certain token distribution should not exist.
These situations would lead to Negative Application Conditions (NACs) in the graph conditions. 

For example, the token distribution shown in \autoref{fig:shippedTwiceProposition} defines a process snapshot with two tokens at activity \textit{Ship goods}.
A modeler could use this atomic proposition to check if the activity \textit{Ship goods} is executed twice by creating an appropriate CTL property.
Shipping goods twice but only receiving one payment during an order-handling process would be a critical error for a business.
The order handling process in \autoref{fig:shippedTwiceProposition} is taken from \cite{ruckerPracticalProcessAutomation2021} but changed to contain a modeling error.
It contains an exclusive gateway instead of a parallel gateway.
The modeling error could lead to shipping goods twice if the process is not corrected before deployment.

\begin{figure}[ht]
    \centering
    \includegraphics[width=0.55\textwidth]{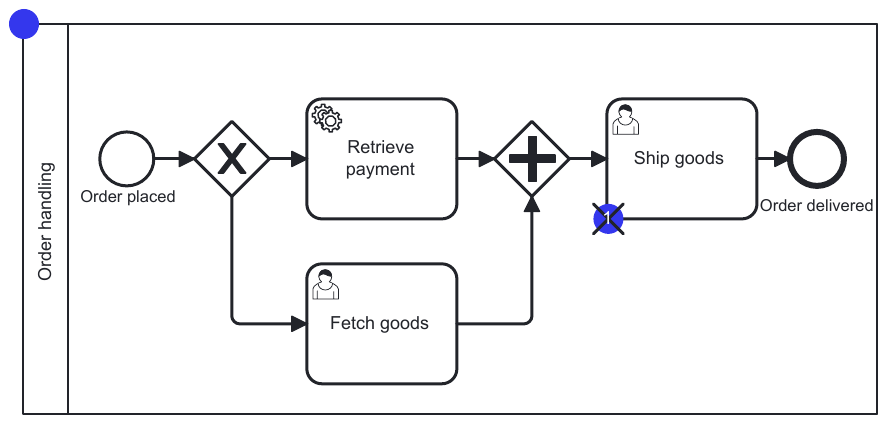}
    \caption{Atomic proposition \textit{noShipment} defining no ongoing shipping of goods.}
    \label{fig:noShipment}
\end{figure}

Another proposition for the same process with a different error is shown in \autoref{fig:noShipment}.
The proposition \textit{noShipment} defines that the activity \textit{Ship goods} should not run (has no token).
\enquote{Has no token} is depicted by crossing out the token symbol and represents an extension of our concrete syntax introduced for defining propositions.
This proposition can be used to define a CTL property to check if shipping always occurs.
In this case, the error in the order handling process prevents shipping from occurring.
The GT systems for both variants of the order handling process containing the propositions can be found in \cite{timkrauterLMCS2024Artifacts2023}.

Using an atomic proposition editor based on the BPMN concrete syntax, modelers do not need extensive knowledge about the GT-based execution semantics. 
Although the expressiveness is not as powerful as in the GT-based execution semantics in Groove---e.g., one can use nested rules with quantification in graph conditions---we favor \textit{simplicity} over \textit{expressiveness}.
In addition, we attempt to stay as independent as possible from the framework and tools used for the execution semantics (see the right part in \autoref{fig:approach}).

Finally, the modeler must still know temporal logic to specify custom properties.
To combat this problem, we added temporal logic templates to our tool to generate commonly occurring propositions without knowledge about temporal logic.
The next section about the BPMN Analyzer discusses this feature in detail.
A domain-specific property language for BPMN would further lessen the knowledge required from the modeler \cite{meyersProMoBoxFrameworkGenerating2014}.

\section{BPMN Analyzer} \label{sec:impl}

\begin{figure}[ht]
    \centering
    \frame{\includegraphics[width=.75\textwidth]{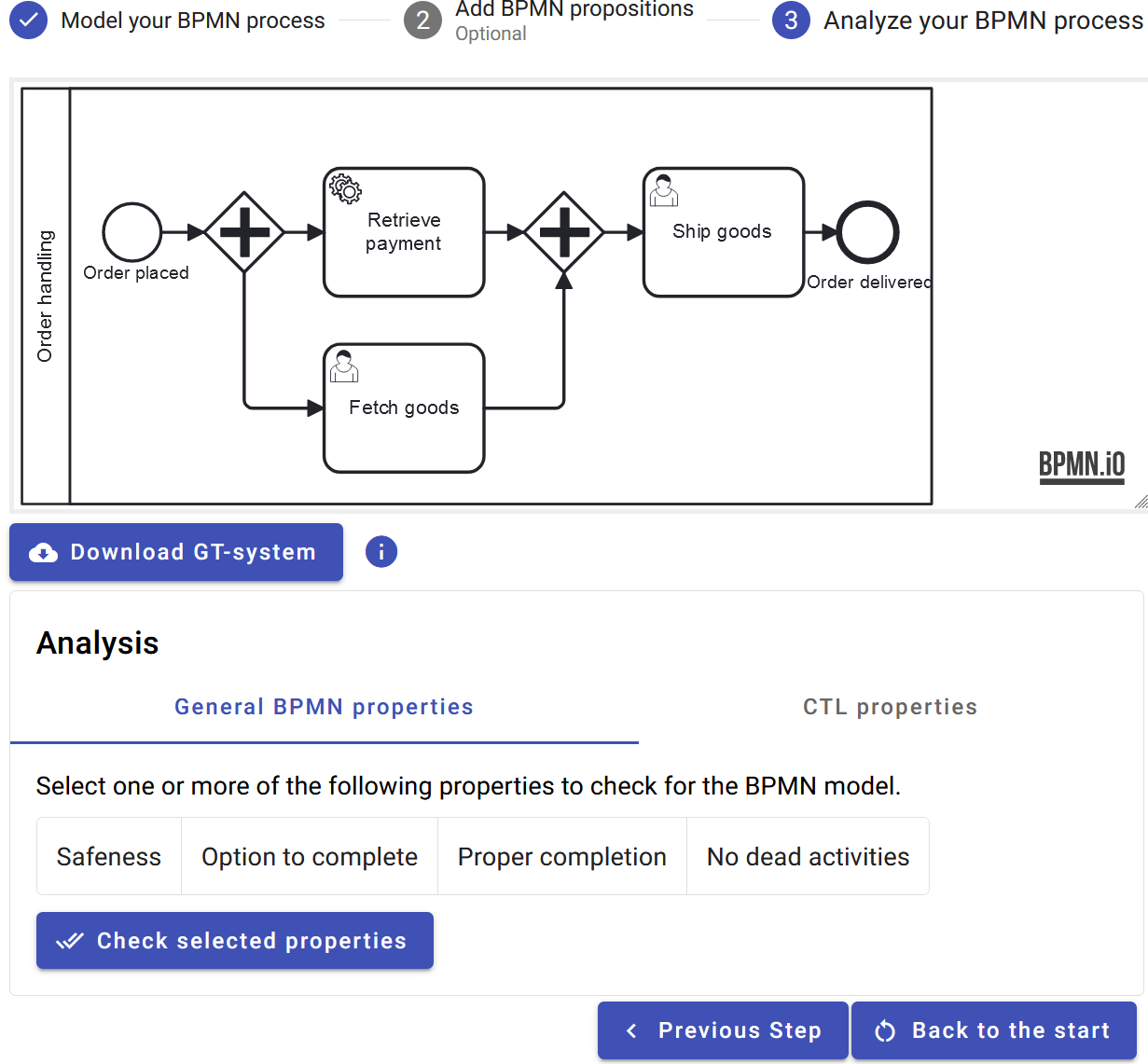}}
    \caption{Screenshot of the \textit{analysis step} in the BPMN Analyzer}
    \label{fig:impl_step1}
\end{figure}

Our approach is implemented as a web-based tool called \textit{BPMN Analyzer}, which is open-source, publicly available, and does not require any installation \cite{timkrauterLMCS2024Artifacts2023, krauterFormalizationAnalysisBPMN2023}.
A demonstration of the BPMN Analyzer is available online\footnote{\url{https://youtu.be/MxXbNUl6IjE}}.
\autoref{fig:impl_step1} depicts a screenshot of the BPMN Analyzer.
We use the order handling process from \cite{ruckerPracticalProcessAutomation2021} as an example.
It is the same BPMN model as in \autoref{fig:shippedTwiceProposition} and \autoref{fig:noShipment} but without modeling errors.

The modeler can create or upload a BPMN model, which can then be verified using either general BPMN properties or custom properties formulated in CTL.
BPMN Analyzer generates a GT system for the supplied BPMN model and runs model checking against the specified properties in Groove \cite{kastenbergModelCheckingDynamic2006, rensinkExplicitStateModel2008}. 
We have created a comprehensive test suite \cite{timkrauterLMCS2024Artifacts2023}, which verifies that rules are generated as defined by the rule generation templates in the previous section.
The test suite covers over 90\% of our source code.

The BPMN Analyzer interface is structured into three steps that guide the user transparently through the modeling and analysis process.

\begin{enumerate}
  \item The \textbf{Modeling} step lets users upload or define the BPMN model.
  We utilize a properties panel in the modeling step so that IDs of BPMN elements can be viewed and edited directly in the model editor.
  This allows for better traceability between BPMN elements and generated GT rules if a user inspects the GT system.
  
  \item The \textbf{BPMN Propositions} step contains our custom \textit{Token Editor} and is shown in \autoref{fig:impl_step2}.
  In this step, users may create atomic propositions, which can be used as ingredients in the custom CTL properties in the analysis step.
  In \autoref{fig:impl_step2}, the user is editing one of two created propositions.
  Users who are only interested in general BPMN properties may skip this step.
  Atomic propositions are created using the concrete syntax detailed in \autoref{subsec:customProperties}, implemented in our \textit{Token Editor}.
  As mentioned, users attach tokens and process snapshots to the BPMN model created in the modeling step to create a proposition.

\begin{figure}[ht]
    \centering
    \frame{\includegraphics[width=0.9\textwidth]{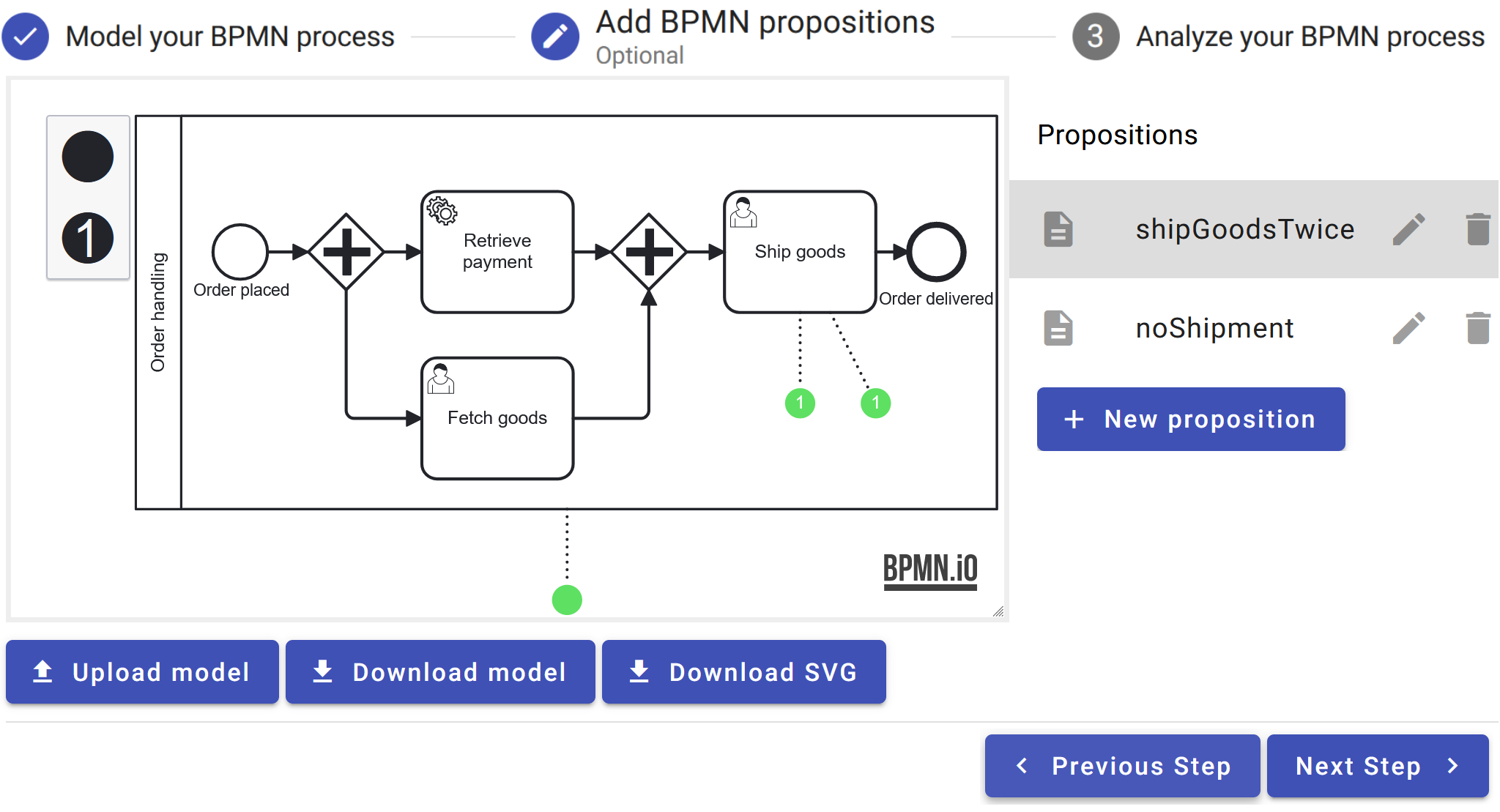}}
    \caption{Screenshot of the \textit{propositions step} in the BPMN Analyzer}
    \label{fig:impl_step2}
\end{figure}
  
  \item In the \textbf{Analysis} step, users may check the general BPMN properties and build custom CTL properties using the atomic propositions.
  The properties builder utilizes the textual CTL syntax implemented in Groove to specify custom properties using the atomic propositions from the previous step, see \autoref{fig:impl_step3_ctl}.
  The two atomic propositions created in the previous step (see \autoref{fig:impl_step2}) are available to the user.
  The CTL properties builder comes with CTL templates to facilitate commonly occurring CTL properties.
  These templates allow users to check whether a state (described by an atomic proposition) can be reached or is never reached (see \autoref{fig:impl_step3_ctl}).
  Thus, simple safety and liveness properties can be checked using these templates.
  Model-checking experts in an organization can define more templates in the future and share them for reuse.
  Users who download the generated GT system may inspect and edit the graph conditions generated from the atomic propositions; they may also specify more properties and check them using Groove.

  \begin{figure}[ht]
      \centering
      \frame{\includegraphics[width=.6\textwidth]{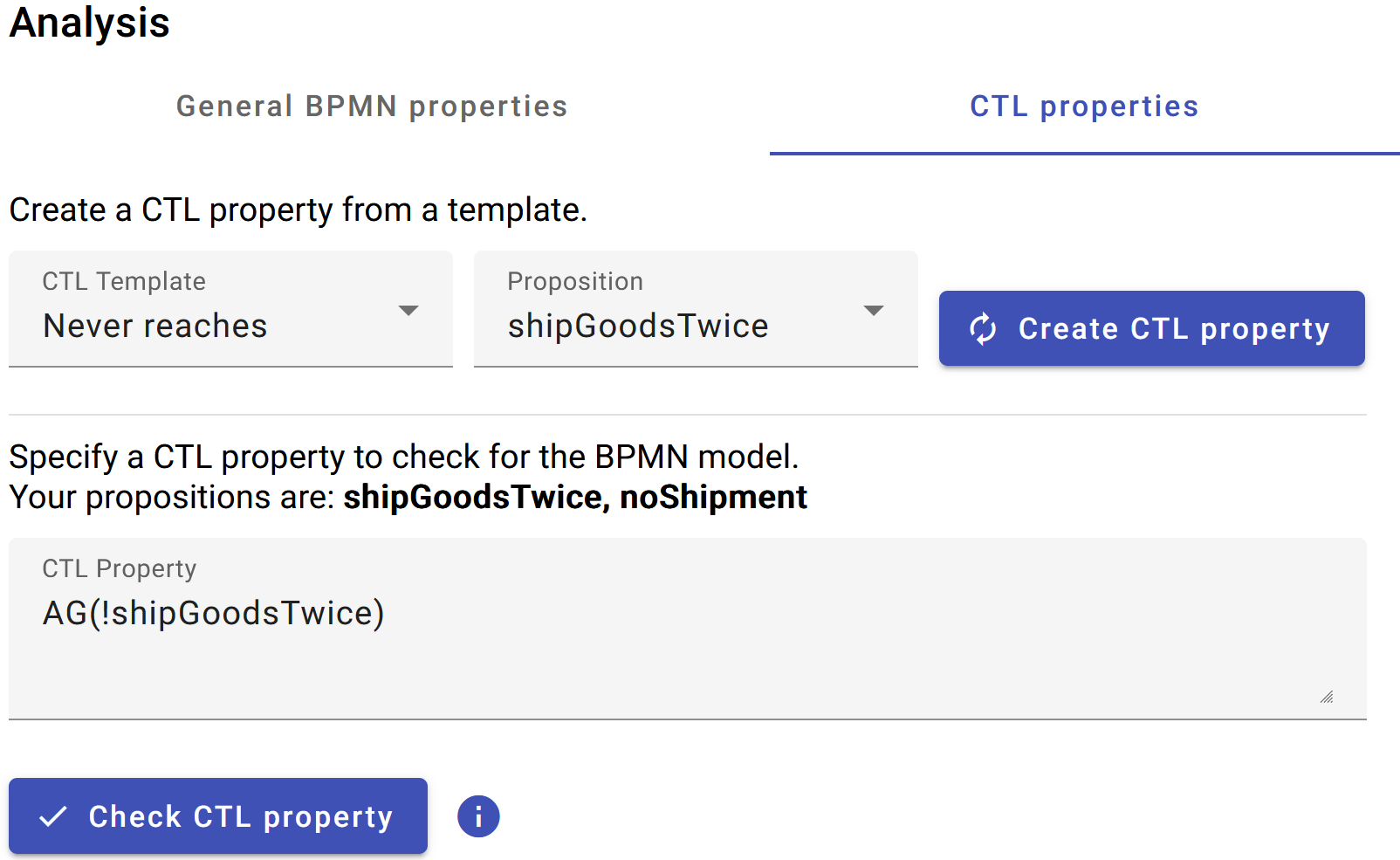}}
      \caption{CTL properties builder in the \textit{analysis step} of the BPMN Analyzer}
      \label{fig:impl_step3_ctl}
  \end{figure}
  
\end{enumerate}

\subsection{Reusable libraries}
In addition to the BPMN Analyzer, we published multiple parts of our application as libraries that can be reused seamlessly by other researchers and practitioners.
The BPMN metamodel extension needed to define atomic propositions is published as an npm module (\textit{token-bpmn-moddle}) \cite{timkrauterLMCS2024Artifacts2023}.
Npm is the default package manager for the JavaScript programming language.
In addition, the Token Editor to create atomic propositions is also published as an npm module (\textit{token-bpmn}) \cite{timkrauterLMCS2024Artifacts2023}.
Furthermore, we published our \textit{graph-rule-generation} library\footnote{\url{https://mvnrepository.com/artifact/io.github.timKraeuter/graph-rule-generation}} to generate Groove GT systems to the Maven central repository \cite{timkrauterLMCS2024Artifacts2023}.
The Maven central repository is the standard repository for developing JVM-based applications.
We explain each library in detail in the following sections.

\subsubsection{BPMN metamodel extension}
Our implementation \textit{token-bpmn-moddle} \cite{timkrauterLMCS2024Artifacts2023} extends \textit{bpmn-moddle} \cite{camundaservicesgmbhBpmnmoddle2023}, which implements the BPMN specification.
Our extension adds the \textsf{Token} and \textsf{ProcessSnapshot} types from the BPMN execution metamodel shown in \autoref{fig:bpmnExecutionMetamodel} to the BPMN metamodel.
\autoref{lst:extension} shows an example BPMN XML snippet, where a token and process snapshot was added.

\begin{lstlisting}[basicstyle=\smaller\ttfamily, label=lst:extension, language=XML, numbers=left,
    stepnumber=1,xleftmargin=\parindent, caption=XML snippet showing the BPMN metamodel extension (simplified)]
<process id="Process_1">
  <extensionElements>
    <bt:processSnapshot id="ProcessSnapshot_1" />
    <bt:token id="Token_1" processSnapshot="ProcessSnapshot_1"
        elementID="Task_A"/>
  </extensionElements>
  <task id="Task_A" />
</process>
\end{lstlisting}

The library allows one to create tokens and process snapshots and stores them in the BPMN extension elements (lines 2-7 in the XML example in \autoref{lst:extension}).
This is the recommended way to extend the BPMN metamodel \cite{objectmanagementgroupBusinessProcessModel2013}.

\begin{figure}[ht]
    \centering
    \includegraphics[width=.6\textwidth]{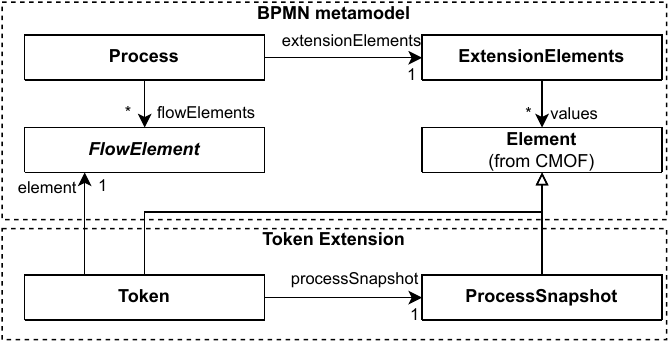}
    \caption{Token Extension of the BPMN metamodel (simplified)}
    \label{fig:token_extension}
\end{figure}

The extension of the BPMN metamodel is realized by letting \textsf{Token} and \textsf{ProcessSnapshot} extend from \textsf{Element}, which is the type of elements of the extension elements, see \autoref{fig:token_extension}.
Each \textsf{Token} points to the \textsf{FlowElement} it is currently positioned at, which can be an \textsf{Activity} or \textsf{SequenceFlow}, see BPMN metamodel in \autoref{fig:bpmnMetamodel}.
In \autoref{lst:extension}, this is realized using the attribute \textsf{elementID}.

The XML model in \autoref{lst:extension} can be used by our HOT to generate atomic propositions for Groove (see \autoref{fig:approach}).
However, following sound model-driven principles and creating an extended metamodel, the XML model could also be used in other applications or for model checking with different tools.

\subsubsection{Token Editor}
The Token Editor implements the concrete syntax for tokens and process snapshots described in \autoref{fig:bpmnExecutionMetamodel}.
Using our Token Editor, a user does not need to write XML, which hides the complexity of extending the BPMN metamodel in the graphical editor.

\autoref{fig:impl_step2} shows the Token Editor embedded in the second step of the BPMN Analyzer.
We simplify the user interface so users can only edit tokens and process snapshots, not the underlying BPMN model.
Process snapshots are automatically assigned distinct colors, and tokens held by a process snapshot have the same color (see concrete syntax in \autoref{fig:bpmnExecutionMetamodel}).
Tokens can be assigned to process snapshots, which changes the token's color to match the snapshot.

Our implementation is based on \textit{bpmn-js} \cite{camundaservicesgmbhBpmnjs2023}, which provides a BPMN rendering toolkit and uses the \textit{token-bpmn-moddle} library described in the previous section to persist and load our models.
Since both implementations are published as libraries, they can be easily reused in other applications.

\subsubsection{Graph rule generation}
The graph-rule-generation library offers various Java classes to generate graphs, GT rules, or entire GT systems, following the \textit{builder pattern} \cite{gammaDesignPatternsElements1995}.
\autoref{lst:grooveRuleBuilder} shows an example code snippet to generate a GT rule using the GT rule builder implemented for Groove.
One could also implement the GT rule builder for a different GT tool than Groove, which would only result in changes in the first two lines of \autoref{lst:grooveRuleBuilder}.

\begin{figure}[ht]
    \centering
    \includegraphics[width=0.4\textwidth]{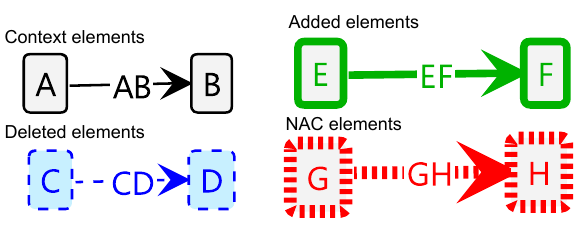}
    \caption{Groove GT rule generated by the code snippet in \autoref{lst:grooveRuleBuilder}}
    \label{fig:grooveRuleBuilderRule}
\end{figure}

\begin{lstlisting}[basicstyle=\smaller\ttfamily, label=lst:grooveRuleBuilder, language=Java,
    stepnumber=1,xleftmargin=\parindent, caption=Code snippet to generate a GT rule using the GT rule builder]
GraphTransformationRuleBuilder ruleBuilder
    = new GrooveRuleBuilder();
// Start a new GT rule with the name "sampleRule"
ruleBuilder.startRule("sampleRule");
// Create context nodes A and B and edge AB from A->B.
GraphNode a = ruleBuilder.contextNode("A");
GraphNode b = ruleBuilder.contextNode("B");
ruleBuilder.contextEdge("AB", a, b);
// Delete nodes C and D and edge CD from C->D.
GraphNode c = ruleBuilder.deleteNode("C");
GraphNode d = ruleBuilder.deleteNode("D");
ruleBuilder.deleteEdge("CD", c, d);
// Add nodes E and F and edge EF from E->F.
GraphNode e = ruleBuilder.addNode("E");
GraphNode f = ruleBuilder.addNode("F");
ruleBuilder.addEdge("EF", e, f);
// Create NAC nodes G and H and edge GH from G->H.
GraphNode g = ruleBuilder.nacNode("G");
GraphNode h = ruleBuilder.nacNode("H");
ruleBuilder.nacEdge("GH", g, h);
// Build the GT rule.
GraphTransformationRule gtRule = ruleBuilder.buildRule();
\end{lstlisting}
Using the rule builder, one can construct a GT rule by defining which nodes and edges should be present (lines 6-9), deleted (lines 10-13), added (lines 14-17), or NACs (lines 18-21), see \autoref{lst:grooveRuleBuilder}.
\autoref{fig:grooveRuleBuilderRule} shows the resulting GT rule specified by \autoref{lst:grooveRuleBuilder} in Groove syntax.

Similarly to the groove rule builder in \autoref{lst:grooveRuleBuilder}, we also provide a builder for constructing graphs to create start graphs of GT systems \cite{timkrauterLMCS2024Artifacts2023}.
Finally, the library provides a builder for GT systems using the graph and GT rule builders.
It also automatically lays out graphs and GT rules using the Eclipse Layout Kernel (ELK).
The Groove UI and the Groove command-line tools can then consume the generated GT systems.

\subsection{Performance testing} \label{subsec:performance}

Model checking is a valuable technique but can fall short when applied in the industry due to insufficient performance \cite{clarkeHandbookModelChecking2018}.
Inadequate performance might have many reasons, most notably large models leading to state space explosion.
We assess the performance of our implementation for two sets of BPMN models.
First, we pick ten BPMN models from the literature, including realistic ones.
Second, we generated ten BPMN models with exponential state space growth to test how our tool deals with increasing complexity.
We explain each benchmark and its results in the following subsections and provide all necessary information and artifacts to reproduce them in \cite{timkrauterLMCS2024Artifacts2023}

To calculate the average runtime, we use the hyperfine benchmarking tool \cite{peterHyperfine2023} (version 1.18.0), which runs the HOT/state space exploration for each BPMN model ten or more times.
The experiment was run on Windows 11 (AMD Ryzen 7700X processor, 32 GB RAM) using Groove version 6.1.0 \cite{timkrauterLMCS2024Artifacts2023}.

\subsubsection{BPMN models from the literature}

We randomly picked ten different BPMN models from \cite{houhouFirstOrderLogicVerification2022} to assess the performance of our implementation.
The models include realistic BPMN models (e.g., 001, 002, and 020) \cite{houhouFirstOrderLogicVerification2022}.

First, we ran our HOT for the BPMN models.
The HOT takes approximately half a second to generate a GT system for each model.
Thus, the generation of the GT systems for these models is fast enough.
In addition, \autoref{table:stateSpaceBenchmark} states how many GT rules are generated for each BPMN model. 

Second, we ran a full state exploration using the resulting ten GT systems, see \autoref{table:stateSpaceBenchmark} (runtime only includes state space exploration).
The exploration takes around one second for most of the models.
Only model \textit{020} needs nearly two seconds due to its larger state space.
Furthermore, up to one second is spent on startup, not model checking.
For example, Groove reports only 722 ms for state space exploration for model \textit{020}.
\begin{table}[ht]
\centering
\caption{Results for a full state space exploration of realistic models}

\begin{tabular}{ c  c  c  c  c  c  c }
 \toprule
 BPMN model & Processes & Nodes (gw.) & GT Rules & States & Transitions & Runtime \\
 \cmidrule(lr){1-3}
 \cmidrule(lr){4-4}
 \cmidrule(lr){5-7}
 001 & 2 & 17 (2) & 26 & 68 & 118 & $\sim$ 1.00 s \\

 002 & 2 & 16 (2) & 24 & 62 & 108 & $\sim$ 0.97 s \\

 007 & 1 & 8 (2) & 14 & 45 & 81 & $\sim$ 0.92 s \\

 008 & 1 & 11 (2) & 17 & 49 & 85 & $\sim$ 0.93 s \\

 009 & 1 & 12 (2) & 17 & 137 & 308 & $\sim$ 1.01 s \\

 010 & 1 & 15 (2) & 20 & 162 & 357 & $\sim$ 1.04 s \\

 011 & 1 & 15 (2) & 20 & 44 & 69 & $\sim$ 0.97 s \\

 015 & 1 & 14 (2) & 20 & 53 & 86 & $\sim$ 0.95 s \\

 016 & 1 & 14 (2) & 19 & 44 & 68 & $\sim$ 0.94 s \\

 020 & 1 & 39 (6) & 59 & 3060 & 8584 & $\sim$ 1.75 s \\
 \bottomrule
\end{tabular}
\label{table:stateSpaceBenchmark}
\end{table}

\subsubsection{BPMN models with increasing complexity}

\begin{figure}[ht]
    \centering
    \includegraphics[width=0.45\textwidth]{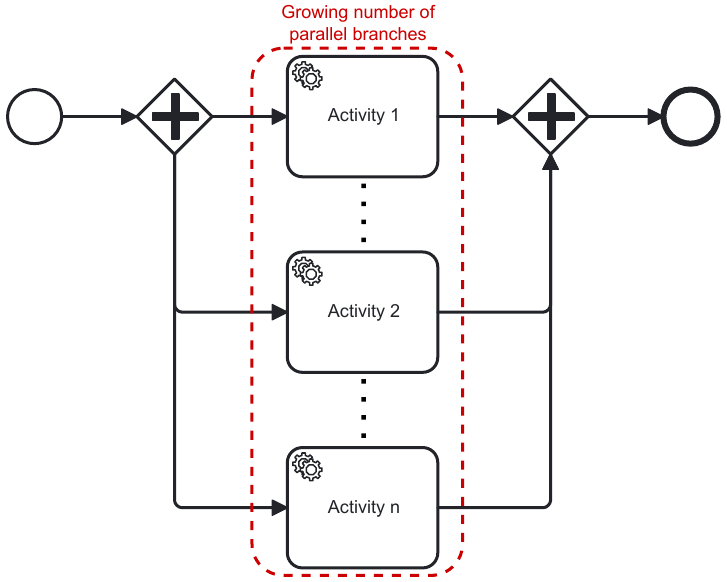}
    \caption{BPMN model generation with an increasing number of parallel branches}
    \label{fig:increasingParallelBranches}
\end{figure}

To increase the state space complexity, we generate BPMN models with a growing number of parallel branches, similar to~\cite{corradiniFormalApproachAnalysis2021}.
\autoref{fig:increasingParallelBranches} shows our schema to generate models.
The possible interleavings of executing activities in parallel lead to an exponential increase in the state space.
Concretely, we generated ten BPMN models with one to ten parallel branches containing one activity \cite{timkrauterLMCS2024Artifacts2023}.
We use these models to benchmark our implementation.

First, we ran our HOT for the BPMN models.
The HOT takes less than a second to generate a GT system for each model.
Thus, the generation of the GT systems for these models is fast enough.

Second, we ran a full state exploration using the resulting ten GT systems, see \autoref{table:parallelBranchesStateSpaceBenchmark} (runtime only includes state space exploration).
The models take one to nine seconds to explore.
One can see an exponential increase in runtime due to the exponential increase in state space complexity.

\begin{table}[ht]
\centering
\caption{Results for a full state space exploration of models with increasing complexity}
\begin{tabular}{ c  c  c  c  c  c }
 \toprule
 Branches & Nodes (gw.) & GT Rules & States & Transitions & Runtime \\
\cmidrule(lr){1-2}
\cmidrule(lr){3-3}
\cmidrule(lr){4-6}
 1 & 5 (2) & 9 & 7 & 7 & $\sim$ 0.87 s \\

 2 & 6 (2) & 11 & 13 & 17 & $\sim$ 0.86 s \\

 3 & 7 (2) & 13 & 31 & 59 & $\sim$ 0.88 s \\

 4 & 8 (2) & 15 & 85 & 221 & $\sim$ 0.95 s \\

 5 & 9 (2) & 17 & 247 & 815 & $\sim$ 1.03 s \\

 6 & 10 (2) & 19 & 733 & 2921 & $\sim$ 1.15 s \\

 7 & 11 (2) & 21 & 2119 & 10.211 & $\sim$ 1.49 s \\

 8 & 12 (2) & 23 & 6.565 & 34.997 & $\sim$ 2.13 s \\

 9 & 13 (2) & 25 & 19.687 & 118.103 & $\sim$ 3.85 s \\

 10 & 14 (2) & 27 & 59.053 & 393.665 & $\sim$ 9.16 s \\
 \bottomrule
\end{tabular}
\label{table:parallelBranchesStateSpaceBenchmark}
\end{table}

We conclude that our approach is sufficiently fast for models of average size and complexity.
In the next section, we test the scalability of our approach when models increase in size.
Furthermore, we discuss potential performance and scalability improvements.
However, a comprehensive benchmark, including a detailed comparison to other tools, is left for future work.

\subsection{Scalability testing} \label{subsec:scalability}

In this section, we test the scalability of our approach by applying it to 300 heterogeneous BPMN models with increasing model sizes.

\subsubsection{Setup}

\begin{figure}[ht]
    \centering
    \includegraphics[width=0.7\textwidth]{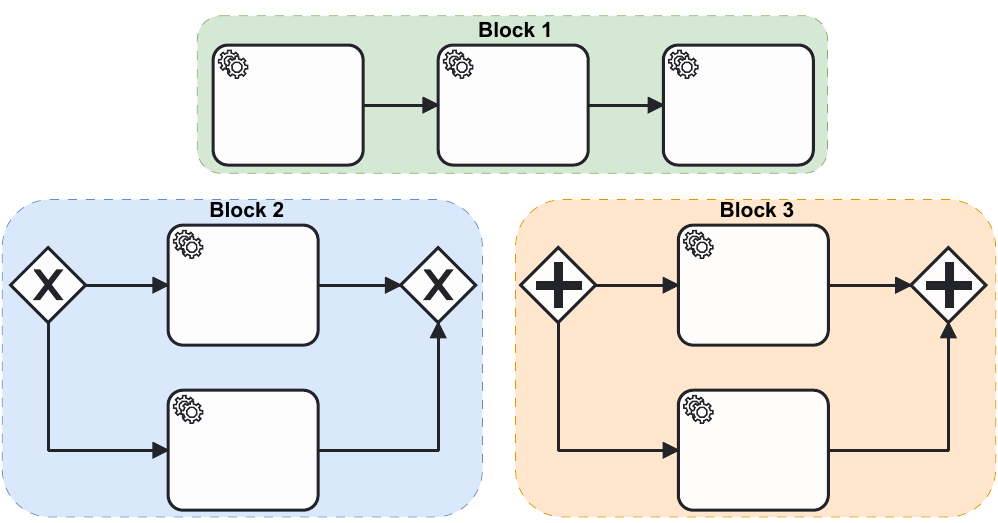}
    \caption{The three different blocks used for BPMN model generation}
    \label{fig:blocks}
\end{figure}

We generated 300 BPMN models to test the scalability of our approach.
We used the following strategy to include different BPMN elements in the models.
We generated the models incrementally, increasing the number of \textit{blocks} they contain.
Thus, model one contains one block, model two contains two blocks, and so forth until the last model contains 300 blocks.
A block is defined as one of the three BPMN model parts shown in \autoref{fig:blocks}.
During the generation, we alternate between the three different blocks.

\begin{figure}[ht]
    \centering
    \includegraphics[width=0.9\textwidth]{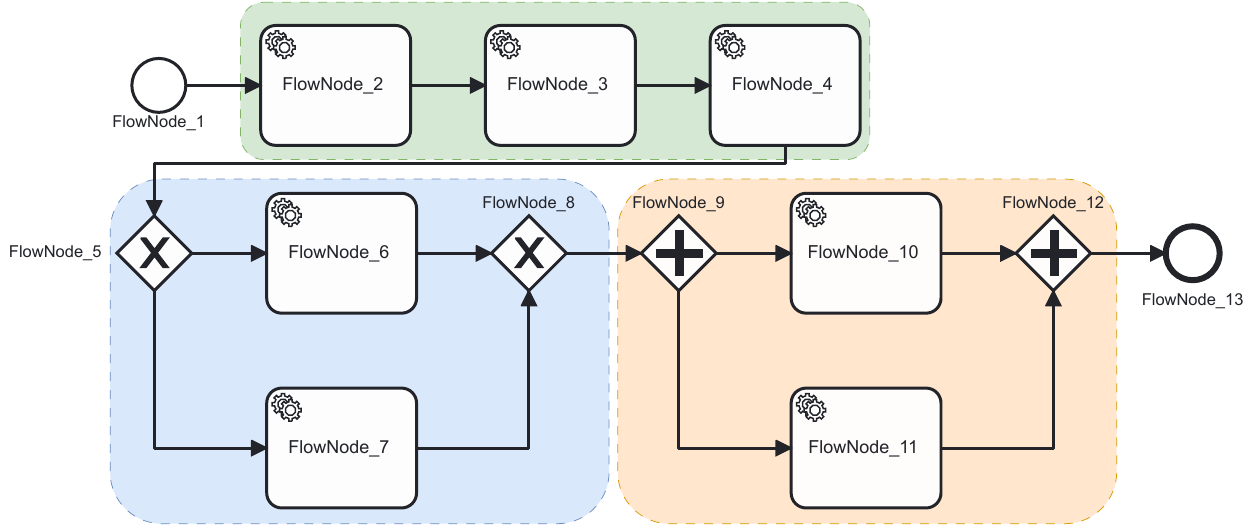}
    \caption{A generated BPMN model with three blocks}
    \label{fig:threeBlockModel}
\end{figure}

For example, the BPMN model with three blocks is depicted in \autoref{fig:threeBlockModel}.
Blocks two and three are shown in a new line for better visualization.
However, the generated models are expanding horizontally in one line.
We then repeat adding one block at a time for each new model until we reach 300 models.

\autoref{table:bpmnModelCharacteristics} states the characteristics of the generated BPMN models, such as the number of gateways, flow nodes, and sequence flows.
One can deduce from the table that adding fifty blocks adds around 400 BPMN elements to a model.
All models, including their characteristics and how to generate them, can be found in \cite{timkrauterLMCS2024Artifacts2023}.
Our BPMN model generation uses the camunda BPMN model API \cite{camundaservicesgmbhCamundaBPMNModel2023}.

BPMN models in practice tend to be much smaller since large models are usually divided into smaller submodels \cite{fahlandAnalysisDemandInstantaneous2011}, i.e., subprocesses, to ensure they are understandable by modelers.
Each of these subprocesses can then be analyzed independently.
From our experience and referring to other studies \cite{fahlandAnalysisDemandInstantaneous2011}, this best practice leads to models with less than 400 total elements (comparable to less than 50 blocks in \autoref{table:bpmnModelCharacteristics}).
We ran our scalability test for models with up to 300 blocks since we wanted enough data to see trends in the average runtime.
We did not go beyond 300 blocks since the whole test should still run in a reasonable time. 

\begin{table}[ht]
\centering
\caption{Characteristics of the generated BPMN models}
\begin{tabular}{ c  c  c  c  c }
 \toprule
 BPMN model / Blocks & Gateways & Flow nodes & Sequence flows & Total elements \\
\cmidrule(lr){1-4}
\cmidrule(lr){5-5}
 1 & 0 & 5 & 4 & 9 \\

 50 & 66 & 185 & 217 & 402 \\

 100 & 132 & 368 & 433 & 801 \\

 150 & 200 & 552 & 651 & 1203 \\

 200 & 266 & 735 & 867 & 1602 \\

 250 & 332 & 918 & 1083 & 2001 \\

 300 & 400 & 1102 & 1301 & 2403 \\
 \bottomrule
\end{tabular}
\label{table:bpmnModelCharacteristics}
\end{table}

\subsubsection{Results}

\begin{figure}[ht]
    \centering
    \includegraphics[width=0.8\textwidth]{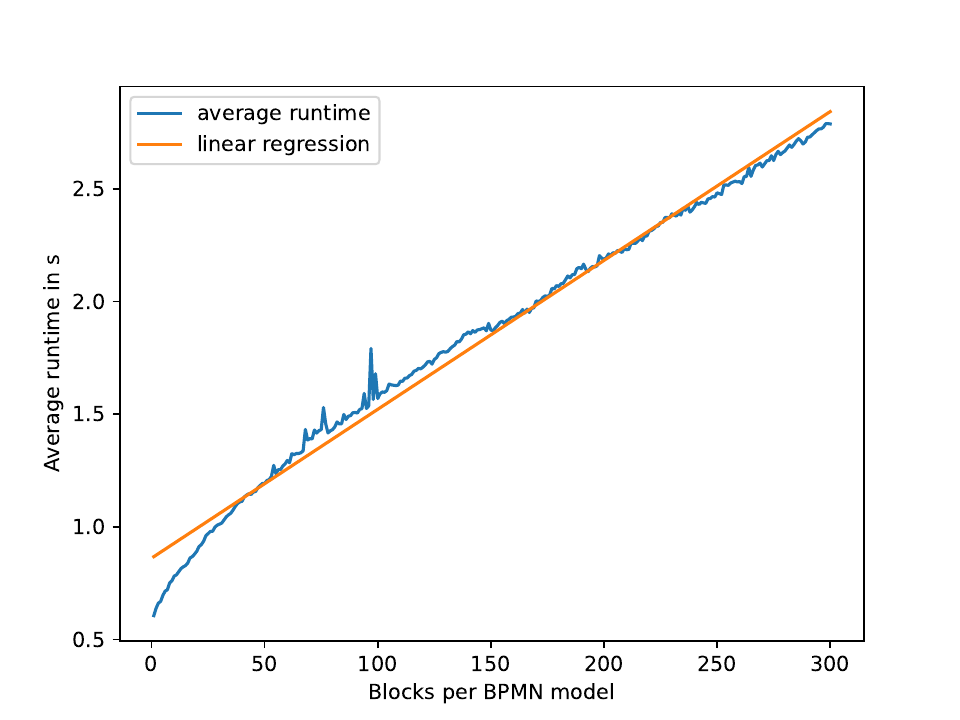}
    \caption{Scalability testing result of the GT system generation}
    \label{fig:hotScalability}
\end{figure}

\autoref{fig:hotScalability} depicts the results of benchmarking our HOT with the generated BPMN models.
It shows the average runtime of five runs for transforming each BPMN model into a GT system using our HOT.
We used the same machine and setup as discussed for our performance experiments in \autoref{sec:impl}.

The average HOT runtimes data fits the linear regression shown in \autoref{fig:hotScalability} well.
This makes sense since the HOT algorithm has \textit{linear runtime complexity} because it iterates over all flow nodes of a BPMN model to generate GT rules.
We conclude that the HOT is fast enough (around one second or less) for models of reasonable size (50 blocks or less).

\begin{figure}[ht]
    \centering
    \includegraphics[width=0.8\textwidth]{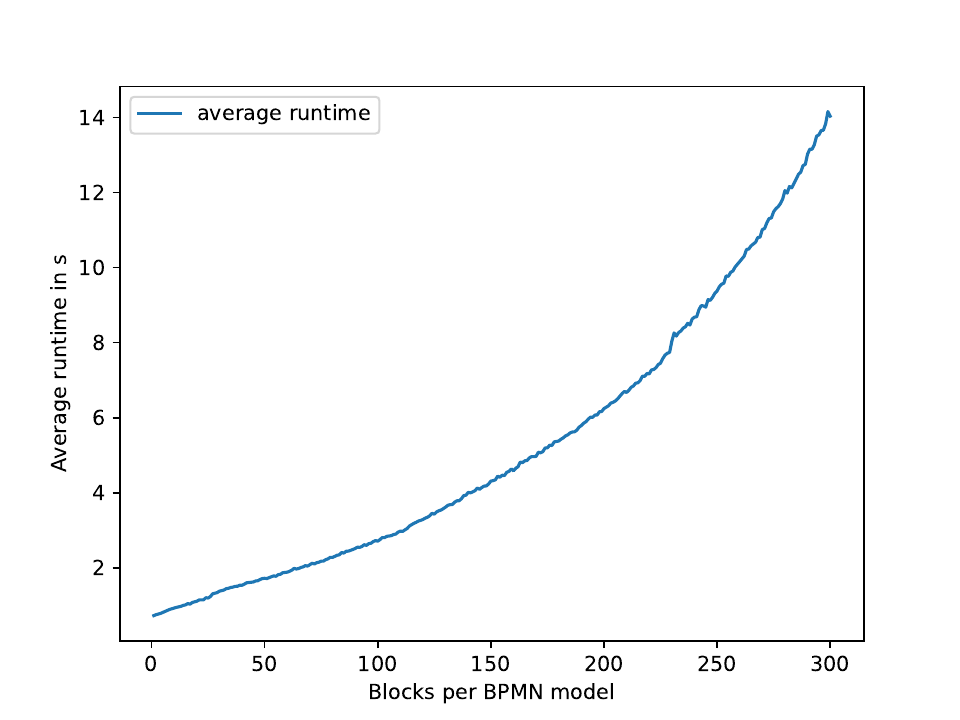}
    \caption{Scalability testing result of the state space generation in Groove}
    \label{fig:stateSpaceScalability}
\end{figure}

\autoref{fig:stateSpaceScalability} depicts the results of benchmarking the state space generation in Groove for the GT systems obtained by our HOT.
It shows the average runtime of five runs, calculated by hyperfine \cite{peterHyperfine2023}.

The increase in the runtime of the state space generation looks worse than linear.
However, models of reasonable size (50 blocks or less) are handled in less than two seconds.
To summarize, using our approach, we can conservatively estimate that these models can be checked (HOT followed by a full state space generation) in around three seconds or less.
In addition, full-state space exploration might not be needed if \textit{on-the-fly} model checking is used \cite{clarkeHandbookModelChecking2018}.

\textit{On-the-fly} model checking allows temporal properties to be checked incrementally while the state space is constructed.
If a property violation is detected, there is no need to further complete the construction of the state space.
This can considerably reduce the time and memory required for verification \cite{clarkeHandbookModelChecking2018}.

Plenty of optimization potential exists, starting from the HOT and ending with the state space generation in Groove.
Currently, neither our HOT nor Groove are specifically optimized for performance with this use case in mind.
Regarding the HOT, multiple optimizations come to mind.
First, one can parallelize the generation of GT rules since each rule is independent.
Second, one can change the rule-generation templates to reduce the number of generated rules and the state space.
For example, two rules are currently generated to represent starting and finishing an activity (see \autoref{fig:activityTemplates}), which fits the description in the BPMN specification.
However, one could instead generate only one rule, which represents the whole execution of an activity.
Thus, one less rule is generated, and the intermediate state representing the activity executing is no longer part of the state space.
If there are many activities, especially when they are executed in parallel, this can lead to large reductions in the state space.
However, a less granular state space could prohibit checking certain properties.
A trade-off exists between staying close to the BPMN execution specification and overall runtime (HOT and state space generation).

Groove is a powerful tool with good out-of-the-box performance.
However, there is still optimization potential.
First, GT rules should not be written to and read from the disk to interact with Groove.
Each GT rule is saved to a new file, and the number of generated GT rules increases with the size of a BPMN model.
Integrating our HOT and Groove more tightly can eliminate these costly I/O operations since the generated rules can stay in the main memory.
Second, \textit{partial order reduction} methods could greatly reduce the time and space required for model checking \cite{clarkeHandbookModelChecking2018}.
Third, Groove could use \textit{on-the-fly} model checking as mentioned by the Groove authors \cite{kastenbergModelCheckingDynamic2006}.
If one combines verification and state space exploration and finds a counterexample, there is no need to continue the state space generation \cite{kastenbergModelCheckingDynamic2006,clarkeHandbookModelChecking2018}.

\section{Related work} \label{sec:relatedWork}
The most common formalizations of BPMN execution semantics use Petri Nets.
For example, \cite{dijkmanSemanticsAnalysisBusiness2008} formalizes a subset of BPMN elements by defining a mapping to Petri Nets conceptually close to our HOT-based formalization.
Especially Petri net model checkers such as LoLA show great performance when analysing business process models \cite{fahlandAnalysisDemandInstantaneous2011} but supporting the same variety of BPMN elements can be problematic.
Encoding basic BPMN modeling elements into Petri Nets is generally straightforward, but for some advanced elements, it can be complicated to define \cite{hofstedeWorkflowPatternsExpressive2002}.
For example, representing \textit{Termination End Events} and \textit{Interrupting Boundary Events}, which interrupt a running process, is usually unsupported because of the complexity of managing the non-local propagation of tokens in Petri Nets \cite{corradiniFormalApproachAnalysis2021}.
We solve these situations by using nested graph conditions, for example, to remove all tokens when reaching a \textit{Termination End Event}.
 
A BPMN formalization based on in-place GT rules is given in \cite{vangorpVisualTokenbasedFormalization2013}.
The formalization covers a substantial part of the BPMN specification, including complex concepts such as inclusive gateways and compensation.
In addition, the GT rules are visual and thus can be aligned with the informal description of the execution semantics of BPMN.
A key difference to our approach is that the rules in \cite{vangorpVisualTokenbasedFormalization2013} are general and can be applied to every BPMN model, while we generate specific rules for each BPMN model using our HOT.
Thus, our approach can be seen as a program specialization compared to \cite{vangorpVisualTokenbasedFormalization2013} since we process a concrete BPMN model before its execution.
However, they do \textit{not} support property checking since their goal is only to formalize the BPMN execution semantics.

The tool \textit{BProVe} is based on formal BPMN semantics given in rewriting logic and implemented in the Maude system \cite{corradiniFormalApproachAnalysis2021}.
Using this formal semantics, \textit{BProVe} can verify custom LTL properties and general BPMN properties, such as Safeness and Soundness.
However, \textit{BProVe} only supports the most common BPMN elements, as shown later.
Regarding performance, \cite{corradiniFormalApproachAnalysis2021} describes a timeout (runtime greater than 600 seconds) for state space generation of a model with five parallel branches.
In \autoref{subsec:performance}, we show that our tool takes less than ten seconds for the state space exploration of a model with ten parallel branches.

The verification framework \textsf{fbpmn} uses first-order logic to formalize and check BPMN models \cite{houhouFirstOrderLogicVerification2022}.
This formalization is then realized in the TLA\textsuperscript{+} formal language, which can be model-checked using TLC.
TLC is an explicit state model checker for TLA\textsuperscript{+} specifications.
Like BProVe, \textsf{fbpmn} allows checking general BPMN properties, such as Safeness and Soundness.
Furthermore, \textit{fbpmn} focuses on different communication models besides the standard in the BPMN specification and supports time-related constructs.
In our approach, we currently disregard time-related constructs \cite{duranVerifyingTimedBPMN2017,houhouFirstOrderLogicVerification2022} and data flow \cite{corradiniFormalisingAnimatingMultiple2022,el-saberCMMICMComplianceChecking2015} but rather support more BPMN elements.
Regarding performance, \cite{houhouFirstOrderLogicVerification2022} report 3.66-10.26s on a machine with less powerful hardware compared to our 1-1.75s for the models in section \autoref{subsec:performance}.
Thus, including the previous comparison to BProVe, we conclude that our tool performs well.
However, assessing tool performance is difficult when there are no standardized benchmarks that allow for a direct comparison of results in the same environment.

\autoref{tab:supportedelements} shows which BPMN elements are supported by our and the abovementioned approaches.
Compared to the other approaches, we cover most BPMN elements.
The coverage of BPMN elements significantly impacts how practical each approach is in checking properties in real life \cite{fahlandAnalysisDemandInstantaneous2011}.
In addition, we cover the most important elements found in practice since we come close to the element coverage of popular process orchestration platforms such as Camunda \cite{camundaservicesgmbhBPMNImplementationReference2023}.

The BPMN elements that our approach does not support, compared to Camunda, are transactions, cancel events, and compensation events.
These elements are rather complex, but \cite{vangorpVisualTokenbasedFormalization2013} shows how cancel and compensation events can be formalized.
We plan to support these elements by extending our implementation and test suite in the future.

\begin{table}[htbp]
    \caption{BPMN elements supported by different formalizations (based on \cite{vangorpVisualTokenbasedFormalization2013}).}
    \label{tab:supportedelements}
    \begin{threeparttable}
    \begin{tabular}{l l l l l l}
    \toprule 
      BPMN element/feature & Dijkman & Van Gorp &  Corradini & Houhou & This\\
      & \cite{dijkmanSemanticsAnalysisBusiness2008} & \cite{vangorpVisualTokenbasedFormalization2013} & \cite{corradiniFormalApproachAnalysis2021}&  \cite{houhouFirstOrderLogicVerification2022} & article\\
      \midrule
      \textit{Instantiation and termination}\\
      Start event instantiation & X & X & X & X & X\\
      Exclusive event-based gateway & & X & & & X \\
        \quad instantiation \\
      Parallel event-based gateway & & & & & \\
        \quad instantiation \\
      Receive task instantiation & & & & & X\\
      Normal process completion & X & X & X & X & X\\
      \\
      \textit{Activities}\\
      Activity & X & X & X & X & X\\
      Loop activity & X & X & & &\\
      Multiple instance activity & & & & & \\
      Subprocess & X & X & & X & X\\
      Event subprocess & &  &  &  & X\\
      Transaction & &  &  &  & \\
      Ad-hoc subprocesses & & & & &\\
      \\
      \textit{Gateways}\\
      Parallel gateway & X & X & X & X & X\\
      Exclusive gateway & X & X & X & X & X\\
      Inclusive gateway (split) & X & X & X & X & X\\
      Inclusive gateway (merge) & & X & & X & X\\
      Event-based gateway & & & X\tnote{1} & X & X\\ 
      Complex gateway & & & & &\\
      \\
      \textit{Events} \\
      None Events & X & X & X & X & X\\
      Message events & X & X & X & X & X\\
      Timer Events & & & & X & \\
      Escalation Events & & & & & X\\
      Error Events & X & X & & & X\\
      Cancel Events & & X & & &\\
      Compensation Events & & X & & &\\
      Conditional Events & & & & &\\
      Link Events & & X & & & X\\
      Signal Events & & X & & & X\\
      Multiple Events & &  & & & \\
      Terminate Events & & X & X & X & X\\
      Boundary Events & & X\tnote{2} & & X\tnote{3} & X\\ 
      \bottomrule
    \end{tabular}
    \begin{tablenotes}
        \item[1] Does not support receive tasks after event-based gateways.
        \item[2] Only supports interrupting boundary events on tasks, not subprocesses.
        \item[3] Only supports message and timer events.
    \end{tablenotes}
    \end{threeparttable}
\end{table}

\section{Conclusion \& future work} \label{sec:conclusion}
This article reports two main practical contributions.
First, we conceptualize a new approach utilizing a Higher-Order model Transformation (HOT) to formalize the semantics of behavioral languages.
Our approach moves complexity from the GT rules to the rule templates, which constitute the HOT.
Furthermore, the approach can be applied to other behavioral languages as long as one can define the \textit{state structure} and identify \textit{state-changing elements} of the language.

Second, we apply our approach to BPMN, resulting in a comprehensive formalization regarding element coverage (compared to the literature and industrial process engines) that supports checking behavioral properties.
Furthermore, our contribution is implemented in an open-source web-based tool to make our ideas easily accessible to other researchers and practitioners.
In addition, our performance and scalability testing indicates that the tool can handle most BPMN models found in practice.

Future work targets both of our main contributions.
First, we plan a detailed comparison of our HOT approach with approaches that utilize fixed model-independent rules.
It will be interesting to investigate how the two approaches differ, for example, in runtime during state space generation.
Second, we aim to improve our formalization and the resulting tool in multiple ways.
We intend to extend our formalization to support the remaining few BPMN elements used in practice and want to turn the modeling environment of our tool into an interactive simulation environment.
In addition, we can use this environment to visualize potential counterexamples in cases where behavioral properties are violated.

\section*{Acknowledgment}
\noindent We thank the anonymous reviewers for their valuable comments and helpful suggestions.

\bibliographystyle{alphaurl} 
\bibliography{bib}

\end{document}